\documentclass[11pt,journal,onecolumn,draftcls]{IEEEtran}

\IEEEoverridecommandlockouts

\usepackage{hyperref}
\usepackage{xr}
\usepackage[table]{xcolor}
\usepackage{array}
\usepackage{bbding}
\usepackage{cite}
\usepackage{graphicx}
\usepackage{amssymb}
\usepackage{amsmath}
\usepackage{amsfonts}
\usepackage{algorithmic}
\usepackage{algorithm}
\usepackage{mdwmath}
\usepackage{mdwtab}
\usepackage{eqparbox}
\usepackage{fixltx2e}
\usepackage{stfloats}
\usepackage{dsfont}
\usepackage{mathrsfs}
\usepackage{bbm}
\usepackage{bm}
\usepackage{threeparttable}
\usepackage{multirow}
\usepackage[caption=false,font=footnotesize]{subfig}
\usepackage{centernot}
\usepackage{accents}
\usepackage{cleveref}
\usepackage{lscape}

\newtheorem{theorem}{Theorem}
\newtheorem{assumption}{Assumption}
\newtheorem{lemma}{Lemma}

\newcommand{\qed}{\nobreak \ifvmode \Relax \else
      \ifdim\lastskip<1.5em \hskip-\lastskip
      \hskip1.5em plus0em minus0.5em \fi \nobreak
      \vrule height0.65em width0.65em depth0em\fi}

\def\defeq{\triangleq}

\DeclareMathOperator*{\minimize}{\mathrm{minimize}}

\def\ds{\mathds}
\def\wt{\widetilde}
\def\wh{\widehat}

\def\col{{\mathrm{col}}}

\def\diag{{\mathrm{diag}}}

\def\Tr{{\mathrm{Tr}}}

\def\kron{\otimes}
\def\bkron{\otimes_{b}}
\def\vecm{{\mathrm{vec}}}
\def\unvecm{{\mathrm{unvec}}}

\def\MSD{{\textrm{MSD}}}

\def\c{{{c}}}
\def\d{{\textrm{d}}}
\def\cent{{\textrm{cent}}}
\def\dist{{\textrm{dist}}}

\def\diff{{\textrm{diff}}}

\def\glob{\textrm{glob}}

\def\sync{{\textrm{sync}}}
\def\async{{\textrm{async}}}

\def\E{\mathbb{E}}
\def\F{\mathbb{F}}

\def\Cov{\mathrm{Cov}}

\def\nn{{\nonumber}}

\newcommand{\be}{\begin{equation}}
\newcommand{\ee}{\end{equation}}
\newcommand{\bea}{\begin{eqnarray*}}
\newcommand{\eea}{\end{eqnarray*}}

\makeatletter
\def\Lddots{\mathinner{\mkern1mu\raise17\p@\vbox{\kern17\p@\hbox{.}}\mkern2mu
    \raise8\p@\hbox{.}\mkern2mu\raise\p@\hbox{.}\mkern1mu}}
 \makeatother

\newcommand{\mbbC}{\mathbb{C}}

\newcommand{\mbbR}{\mathbb{R}}
\newcommand{\mbbT}{\mathbb{T}}

\newcommand{\Acal}{\mathcal{A}}
\newcommand{\Bcal}{\mathcal{B}}

\newcommand{\Dcal}{\mathcal{D}}

\newcommand{\Fcal}{\mathcal{F}}

\newcommand{\Hcal}{\mathcal{H}}

\newcommand{\Jcal}{\mathcal{J}}

\newcommand{\Mcal}{\mathcal{M}}
\newcommand{\Ncal}{\mathcal{N}}
\newcommand{\Pcal}{\mathcal{P}}
\newcommand{\Qcal}{\mathcal{Q}}
\newcommand{\Rcal}{\mathcal{R}}

\newcommand{\Tcal}{\mathcal{T}}

\newcommand{\Xcal}{\mathcal{X}}

\def\Nknk{\Ncal_k\backslash\{k\}}

\newcommand{\one}{\ds{1}}

\newcommand{\T}{\mathsf{T}} 
\newcommand{\ubar}{\underaccent{\bar}}

\begin{document}
%
\title{Asynchronous Adaptation and Learning over Networks --- Part III: Comparison Analysis}

\author{Xiaochuan~Zhao,~\IEEEmembership{Student~Member,~IEEE,}
        and Ali~H.~Sayed,~\IEEEmembership{Fellow,~IEEE} \\
\thanks{The authors are with Department of Electrical Engineering, University of California, Los Angeles, CA 90095
Email: xiaochuanzhao@ucla.edu, sayed@ee.ucla.edu.}
\thanks{This work was supported by NSF grants CCF-1011918 and ECCS-1407712. A short and limited early version of this work appeared in the conference proceeding \cite{Zhao12EUSIPCO}. The first two parts of this work are presented in \cite{Zhao13TSPasync1, Zhao13TSPasync2}.}}


\maketitle

\begin{abstract}
In Part II \cite{Zhao13TSPasync2} we carried out a detailed mean-square-error analysis of the performance of asynchronous adaptation and learning over networks under a fairly general model for asynchronous events including random topologies, random link failures, random data arrival times, and agents turning on and off randomly. In this Part III, we compare the performance of synchronous and asynchronous networks. We also compare the performance of decentralized adaptation against centralized stochastic-gradient (batch) solutions. Two interesting conclusions stand out. First, the results establish that the performance of adaptive networks is largely immune to the effect of asynchronous events: the mean and mean-square convergence rates and the asymptotic bias values are not degraded relative to synchronous or centralized implementations. Only the steady-state mean-square-deviation suffers a degradation in the order of $\nu$, which represents the small step-size parameters used for adaptation. Second, the results show that the adaptive distributed network matches the performance of the centralized solution. These conclusions highlight another critical benefit of cooperation by networked agents: cooperation does not only enhance performance in comparison to stand-alone single-agent processing, but it also endows the network with remarkable resilience to various forms of random failure events and is able to deliver performance that is as powerful as batch solutions.
\end{abstract}

\begin{IEEEkeywords}
Distributed optimization, diffusion adaptation, asynchronous behavior, centralized solutions, batch solutions, adaptive networks, dynamic topology, link failures.
\end{IEEEkeywords}

\allowdisplaybreaks

\section{Introduction}
In Part I \cite{Zhao13TSPasync1} we introduced a general model for asynchronous behavior over adaptive networks that allowed for various sources of uncertainties including random topologies, random link failures, random data arrival times, and agents turning on and off randomly. We showed that despite these uncertainties, which could even occur simultaneously, the adaptation process remains mean-square stable for sufficiently small step-sizes. Specifically, we derived condition \eqref{I-eqn:meansquarestabilitycond3} in Part I \cite{Zhao13TSPasync1}, namely,
\be
\label{eqn:meansquarestabilitycond3}
\frac{\bar{\mu}_k^{(2)}}{\bar{\mu}_k^{(1)}} < \frac{\lambda_{k,\min}}{\lambda_{k,\max}^2 + \alpha}
\ee
for all $k$, to ensure that the steady-state individual mean-square-deviations (MSD) satisfies
\be
\label{eqn:individualMSDapproxOnu}
\limsup_{i\rightarrow \infty} \E\,\|w^o - \bm{w}_{k,i}\|^2 = O(\nu)
\ee
for all $k$, where $\bar{\mu}_k^{(m)} \defeq \E[ \bm{\mu}_k(i) ]^m$ denotes the $m$-th moment of the random step-size parameter $\bm{\mu}_k(i)$, $\{ \lambda_{k,\min}, \lambda_{k, \max}, \alpha \}$ are from Assumptions \ref{I-asm:boundedHessian} and \ref{I-asm:gradientnoise} of Part I \cite{Zhao13TSPasync1}, $w^o$ denotes the optimal minimizer, and 
\be
\label{eqn:nudef}
\nu \defeq \max_{k} \frac{\sqrt{\bar{\mu}_k^{(4)}}}{\bar{\mu}_k^{(1)} }
\ee
Note that in Theorem \ref{I-theorem:meansquarestability} from Part I \cite{Zhao13TSPasync1}, we used $\nu_o$ in \eqref{eqn:individualMSDapproxOnu}, where $\nu_o$ is from \eqref{I-eqn:mumaxdef} of Part I \cite{Zhao13TSPasync1}. Since $\nu_o \le \nu$ by \eqref{I-eqn:nuandnuo} from Part I \cite{Zhao13TSPasync1}, we replaced $\nu_o$ with $\nu$ in \eqref{eqn:individualMSDapproxOnu}.

In Part II \cite{Zhao13TSPasync2} we examined the attainable mean-square-error (MSE) performance of the asynchronous network and derived expressions that reveal how close the estimates at the various agents get to the desired optimal solution that is sought by the network. In particular, we showed among other results that under a strengthened condition \eqref{II-eqn:boundstepsize4thorder} from Part II \cite{Zhao13TSPasync2} (relative to condition \eqref{eqn:meansquarestabilitycond3}), namely,
\be
\label{eqn:boundstepsize4thorder}
\frac{\sqrt{\bar{\mu}_k^{(4)}}}{\bar{\mu}_k^{(1)} } < \frac{ \lambda_{k, \min} }{ 3\lambda_{k, \max}^2 + 4\alpha }
\ee
for all $k$, it holds that
\be
\label{eqn:crossMSDapproxOnu2}
\limsup_{i\rightarrow \infty} \E\,\|\bm{w}_{k,i} - \bm{w}_{\ell,i}\|^2 = O(\nu^{1 + \gamma_o'})
\ee
for all $k$ and $\ell$, where $\gamma_o' > 0$ is some constant that is given by \eqref{II-eqn:gammaopdef} of Part II \cite{Zhao13TSPasync2}.

In \eqref{I-eqn:oldnulessnewnu} and \eqref{I-eqn:newboundlessoldbound} from Appendix \ref{I-app:4thorder} of Part I \cite{Zhao13TSPasync1}, we showed that condition \eqref{eqn:boundstepsize4thorder} implies condition \eqref{eqn:meansquarestabilitycond3} so that both results \eqref{eqn:individualMSDapproxOnu} and \eqref{eqn:crossMSDapproxOnu2} hold. Expressions \eqref{eqn:individualMSDapproxOnu} and \eqref{eqn:crossMSDapproxOnu2} show that all agents are able to reach a level of $O(\nu^{1 + \gamma_o'})$ agreement with each other and to get $O(\nu)$ close to $w^o$ in steady-state. These results establish that asynchronous networks can operate in a stable manner under fairly general asynchronous events and, importantly, are able to adapt and learn well. Two important questions remain to be addressed:
\begin{enumerate}
\item Compared with synchronous networks, does the asynchronous behavior degrade performance?
\item How close can the performance of an asynchronous network get to that of a centralized solution?
\end{enumerate}
In this Part III, we therefore compare the performance of synchronous and asynchronous networks. We also compare the performance of distributed solutions against centralized (batch) solutions. The results will show that the performance of adaptive networks are surprisingly immune to the effect of asynchronous uncertainties: the mean and mean-square convergence rates and the asymptotic bias values are not degraded relative to synchronous or centralized implementations. Only the steady-state MSD suffers a degradation of the order of $\nu$. The results also show that an adaptive network always matches the performance of a centralized solution. The main results of this part are summarized in Table \ref{table:summary}, which compares various performance metrics across different implementations. The notation in Table \ref{table:summary} will be explained in the sequel. For now, we simply remark that the results in Table \ref{table:summary} show that the distributed and centralized implementations have almost the same mean-square performance in either the synchronous or asynchronous modes of operation, i.e., the asynchronous distributed implementation approaches the asynchronous centralized implementation, and the synchronous distributed implementation approaches the synchronous centralized implementation. 

We indicated in the introductory remarks of Part I \cite{Zhao13TSPasync1} that studies exist in the literature that examine the performance of distributed strategies in the presence of some forms of asynchronous uncertainties \cite{Tsitsiklis86TAC, Boyd06TIT, Kar09TSP, Srivastava11JSTSP} or changing topologies \cite{Boyd06TIT, Kar08TSP, Kar09TSP, Aysal09Allerton, Aysal09TSP, Kar10TSP, Jakovetic10TSP, Kar11TSP, Srivastava11JSTSP}, albeit for decaying step-sizes. We also explained how the general asynchronous model that we introduced in Part I \cite{Zhao13TSPasync1} covers broader situations of practical interest, including adaptation and learning under constant step-sizes, and how it allows for the simultaneous occurrence of multiple random events from various sources. Still, these earlier studies do not address the two questions posed earlier on how asynchronous networks compare in performance to synchronous networks and to centralized (batch) solutions. If it can be argued that asynchronous networks are still able to deliver performance similar to synchronous implementations where no uncertainty occurs, or similar to batch solutions where all information is aggregated and available for processing in a centralized fashion, then such a conclusion would be of significant practical relevance. The same conclusion would provide a clear theoretical justification for another critical benefit of cooperation by networked agents, namely, that cooperation does not only enhance performance in comparison to stand-alone single-agent processing, as already demonstrated in prior works in the literature (see, e.g., \cite{Sayed13Chapter, Sayed13SPM, Sayed14PROC} and the references therein), but it also endows the network with remarkable resilience to various forms of uncertainties and is still able to deliver performance that is as powerful as batch solutions.

\begin{landscape}
\begin{table}
\renewcommand{\arraystretch}{1.5}
\caption{Comparison of synchronous vs. asynchronous and distributed vs. centralized solutions}
\label{table:summary}
\centering
\begin{threeparttable}
\begin{tabular}{|c||c|c|c|c|}
\hline 
\bfseries {} & \bfseries Synchronous Distributed & \bfseries Asynchronous Distributed & \bfseries Synchronous Centralized & \bfseries Asynchronous Centralized \\ 
\hline 
\hline
\bfseries Algs. & \eqref{eqn:atcsync1} and \eqref{eqn:atcsync2} & \eqref{eqn:atcasync1} and \eqref{eqn:atcasync2} & \eqref{eqn:batchsyn} & \eqref{eqn:batchasyn} \\
\hline
\bfseries Vars. \tnote{a} & $\bm{w}_{i,\sync}^{\diff}$ & $\bm{w}_{i,\async}^{\diff}$ & $\bm{w}_{i,\sync}^{\cent}$ & $\bm{w}_{i,\async}^{\cent}$ \\
\hline
\bfseries Paras. \tnote{b} & $\{\bar{a}_{\ell k}, \bar{\mu}_k\}$ & $\{\bm{a}_{\ell k}(i), \bm{\mu}_k(i)\}$ & $\{\bar{\pi}_{k}, \bar{\mu}_k\}$ & $\{\bm{\pi}_{k}(i), \bm{\mu}_k(i)\}$ \\
\hline
\bfseries Mn. Rate \tnote{c} & $\rho(\bar{\Bcal}) = \rho_o  +  O(\nu^{1+1/N})$ & $\rho(\bar{\Bcal}) = \rho_o  +  O(\nu^{1+1/N})$ & $\rho(\bar{B}) = \rho_o$ & $\rho(\bar{B}) = \rho_o$ \\ 
\hline 
\bfseries M.S. Rate \tnote{d} & $\rho(\Fcal_{\sync}) = \rho_o^2  +  O(\nu^{1+1/N})$ & $\rho(\Fcal_{\async}) = \rho_o^2  +  O(\nu^{1+1/N^2})$ & $\rho(F_{\sync}) = \rho_o^2$ & $\rho(F_{\async}) = \rho_o^2 + O(\nu^2)$ \\ 
\hline 
\bfseries MSD \tnote{e} & $\frac{1}{4}\Tr(H^{-1}   R_{\sync})  +  O(\nu^{1+\gamma_o})$ & $\frac{1}{4}\Tr(H^{-1}   R_{\async})  +  O(\nu^{1+\gamma_o})$ & $\frac{1}{4}\Tr(H^{-1}   R_{\sync})  +  O(\nu^{1+\gamma_o})$ & $\frac{1}{4}\Tr(H^{-1}   R_{\async})  +  O(\nu^{1+\gamma_o})$ \\ 
\hline 
\end{tabular} 

\smallskip

\begin{tablenotes}
\item[a] Variables. The variables for synchronous diffusion strategies are denoted in the table by $\bm{w}_{i,\sync}^{\diff} \defeq \col\{\bm{w}_{1,i,\sync}^{\diff}, \bm{w}_{2,i,\sync}^{\diff}, \dots, \bm{w}_{N,i,\sync}^{\diff}\}$, where $\bm{w}_{k,i,\sync}^{\diff}$ denotes the iterate of agent $k$ at time $i$. The variables for asynchronous diffusion strategies are defined in the same manner.

\item[b] Parameters. The parameters used by the four strategies satisfy:
\begin{enumerate}
\item $\bar{\mu}_k = \E\,[\bm{\mu}_k(i)]$.

\item $c_{\mu,k,\ell} = \E\,[(\bm{\mu}_k(i) - \bar{\mu}_k)(\bm{\mu}_\ell(i) - \bar{\mu}_\ell)]$.

\item $\bar{a}_{\ell k} = \E\,[\bm{a}_{\ell k}(i)]$, $\bar{\pi}_k = \E\,[\bm{\pi}_k(i)]$, and $\bar{\pi}_k = \bar{p}_k$, where $\bar{A} = [\bar{a}_{\ell k}]_{\ell,k=1}^{N}$, $\bar{p} = [\bar{p}_k]_{k=1}^{N}$, $\bar{A}\bar{p} = \bar{p}$, and $\bar{p}^\T\one_N = 1$.

\item $c_{a,\ell k,nm} = \E\,[(\bm{a}_{\ell k}(i) - \bar{a}_{\ell k})(\bm{a}_{n m}(i) - \bar{a}_{n m})]$, $c_{\pi,k,\ell} = \E\,[(\bm{\pi}_k(i) - \bar{\pi}_k)(\bm{\pi}_\ell(i) - \bar{\pi}_\ell)]$, and $C_{\pi} = P_p - \bar{p}\bar{p}^\T$,  where $C_A = [c_{a,\ell k,nm}]_{\ell,k,n,m=1}^{N}$, $C_{\pi} = [c_{\pi,k,\ell}]_{\ell,k=1}^{N}$, $p = \vecm(P_p)$, $(\bar{A} \kron \bar{A} + C_A)p = p$, and $p^\T \one_{N^2} = 1$.
\end{enumerate}

\item[c] Mean convergence rates. The matrices $\{\bar{\Bcal}, \bar{B}\}$ are given by \eqref{II-eqn:bigBmeandef} from Part II \cite{Zhao13TSPasync2} and \eqref{eqn:Bmeandef} in this part. Moreover, $\rho_o \defeq 1 - \lambda_{\min}(H) = 1 - O(\nu)$, where $H$ is given by \eqref{II-eqn:Hdef} from Part II \cite{Zhao13TSPasync2}.

\item[d] Mean-Square convergence rates. The matrices $\{\bar{\Fcal}_\sync, \bar{\Fcal}_\async, F_\sync, F_\async\}$ are given by \eqref{eqn:bigFsyncdef}, \eqref{eqn:bigFasyncdef}, \eqref{eqn:Fsyncdef}, and \eqref{eqn:Fcdef}, respectively.

\item[e] Mean-Square-Deviations. The matrices $\{R_\sync, R_\async\}$ are given by \eqref{eqn:Rsyncdef} and \eqref{eqn:Rcdef}, respectively, and $\gamma_o$ is given by \eqref{II-eqn:gammaodef} of Part II \cite{Zhao13TSPasync2}. Moreover, $R_{\async} - R_{\sync} = O(\nu^2) > 0$.
 
\end{tablenotes}
\end{threeparttable}
\end{table}
\end{landscape}

For the remainder of this part, we continue to use the same symbols, notation, and assumptions from Part I \cite{Zhao13TSPasync1} and Part II \cite{Zhao13TSPasync2}. Moreover, we focus on presenting the main results and their interpretation in the body of the paper, while delaying the technical derivations and arguments to the appendices.

\section{Centralized Batch Solution}
We first describe and examine the centralized (batch) solution. In order to allow for a fair comparison among the various implementations, we assume that the centralized solution is also running a stochastic-gradient approximation algorithm albeit one that has access to the \emph{entire} set of data at each iteration. Obviously, centralized solutions can be more powerful and run more complex optimization procedures. Our purpose is to examine the various implementations under similar algorithmic structures and complexity.

\subsection{Centralized Solution in Two Forms}

We thus consider a scenario where there is a fusion center that regularly collects the data from across the network and is interested in solving the same minimization problem \eqref{I-eqn:globalcost} as in Part I \cite{Zhao13TSPasync1}, namely,
\begin{align}
\label{eqn:globalcost}
\minimize_{w}\;\;J^{\glob}(w)\defeq\sum_{k=1}^{N}J_k(w)
\end{align}
where the cost functions $\{J_k(w)\}$ satisfy Assumptions \ref{I-asm:costfunctions} and \ref{I-asm:boundedHessian} in Part I \cite{Zhao13TSPasync1} and each has a unique minimizer at $w^o \in \mbbC^{M}$. The fusion center seeks the optimal solution $w^o$ of \eqref{eqn:globalcost} by running an \emph{asynchronous} stochastic gradient batch algorithm of the following form (later in \eqref{eqn:batchsyn} we consider a synchronous version of this batch solution):
\be
\label{eqn:batchasyn}
{\bm{w}}_{\c,i} = {\bm{w}}_{\c,i-1} - \sum_{k=1}^{N} \bm{\pi}_{k}(i) \bm{\mu}_k(i) \wh{\nabla_{w^*}J_k}(\bm{w}_{\c,i-1})
\ee
where ${\bm{w}}_{\c,i}$ denotes the iterate at time $i$, the $\{\bm{\pi}_{k}(i)\}$ are nonnegative convex fusion coefficients such that
\be
\label{eqn:convexcondition}
\sum_{k=1}^{N} \bm{\pi}_{k}(i) = 1, \qquad \bm{\pi}_{k}(i) \ge 0
\ee
for all $i\ge0$, and the $\{\bm{\mu}_k(i)\}$ are the random step-sizes. 

We will describe later in \eqref{eqn:batchsyn} the centralized implementation for the synchronous batch solution. In that implementation, all agents transmit their data to the fusion center. In contrast, the implementation in \eqref{eqn:batchasyn} allows the transmission of data from agents to occur in an asynchronous manner. Specifically, we use \emph{random} step-sizes $\{\bm{\mu}_k(i)\}$ in \eqref{eqn:batchsyn} to account for random activity by the agents, which may be caused by random data arrival times or by some power saving strategies that turn agents on and off randomly. We also use \emph{random} fusion coefficients $\{\bm{\pi}_{k}(i)\}$ to model the random status of the communication links connecting the agents to the fusion center. This source of randomness may be caused by random fading effects over the communication channels or by random data feeding/fetching strategies. Therefore, the implementation in \eqref{eqn:batchasyn} is able to accommodate various forms of asynchronous events arising from practical scenarios, and is a useful extension of the classical batch solution in \eqref{eqn:batchsyn}.

It is worth noting that the centralized (batch) algorithm \eqref{eqn:batchasyn} admits a decentralized, though not fully-distributed, implementation of the following form:
\begin{subequations}
\begin{alignat}{2}
\label{eqn:batchasyn1}
{\bm{\psi}}_{k,i} & = {\bm{w}}_{\c,i-1} - \bm{\mu}_k(i) \wh{\nabla_{w^*}J_k}(\bm{w}_{\c,i-1}) & \;\; & \mbox{(adaptation)}\\
\label{eqn:batchasyn2}
{\bm{w}}_{\c,i} & = \sum_{k=1}^{N} \bm{\pi}_{k}(i) \, {\bm{\psi}}_{k,i} & \;\; & \mbox{(fusion)}
\end{alignat}
\end{subequations}
In this description, each agent $k$ uses the local gradient data to calculate the intermediate iterate ${\bm{\psi}}_{k,i}$ and feeds its value to a fusion center; the fusion center fuses all intermediate updates $\{{\bm{\psi}}_{k,i}\}$ according to \eqref{eqn:batchasyn2} to obtain ${\bm{w}}_{\c,i}$ and then forwards the results to all agents. This process repeats itself at every iteration. Implementation \eqref{eqn:batchasyn1}--\eqref{eqn:batchasyn2} is not fully distributed because, for example, all agents require knowledge of the same global iterate ${\bm{w}}_{\c,i}$ to perform the adaptation step \eqref{eqn:batchasyn1}. Since the one-step centralized implementation \eqref{eqn:batchasyn} and the two-step equivalent  \eqref{eqn:batchasyn1}--\eqref{eqn:batchasyn2} represent the same algorithm, we shall use them interchangeably to facilitate the analysis whenever necessary. One advantage of the decentralized representation \eqref{eqn:batchasyn1}--\eqref{eqn:batchasyn2} is that it can be viewed as a distributed solution over \emph{fully}-connected networks \cite{Zhao13ICASSP}.

\subsection{Gradient Noise and Asynchronous Models}
\label{subsec:model}
We assume that the approximate gradient vector $\wh{\nabla_{w^*}J_k}(\bm{w}_{\c,i-1})$ in \eqref{eqn:batchasyn} follows the same model described by \eqref{I-eqn:linearperturbationmodel} in Part I \cite{Zhao13TSPasync1}, namely,
\be
\label{eqn:gradientmodel}
\wh{\nabla_{w^*}J_k}(\bm{w}_{\c,i-1}) = \nabla_{w^*}J_k(\bm{w}_{\c,i-1}) + \bm{v}_{k,i}(\bm{w}_{\c,i-1})
\ee
where the first term on the RHS is the true gradient and the second term models the uncertainty about the true gradient. We continue to assume that the gradient noise $\bm{v}_{k,i}(\bm{w}_{\c,i-1})$ satisfies Assumption \ref{II-asm:gradientnoisestronger} from Part II \cite{Zhao13TSPasync2}.

From Assumption \ref{II-asm:gradientnoisestronger} of Part II \cite{Zhao13TSPasync2}, the conditional moments of $\ubar{\bm{v}}_{k,i}(\bm{w}_{k,i-1})$ satisfy
\begin{align}
\label{eqn:gradientnoisemean}
\E[\ubar{\bm{v}}_{k,i}(\bm{w}_{k,i-1}) | \F_{i-1} ] & = 0 \\
\label{eqn:gradientnoise4thmoment}
\E[\|\ubar{\bm{v}}_{k,i}(\bm{w}_{k,i-1})\|^4 | \F_{i-1} ] & \le \alpha^2 \|\ubar{w}^o - \ubar{\bm{w}}_{k,i-1}\|^4 + 4 \sigma_v^4
\end{align}
where a factor of 4 appeared due to the transform $\ubar{\mbbT}(\cdot)$ from \eqref{I-eqn:ubardef} of Part I \cite{Zhao13TSPasync1}.

To facilitate the comparison in the sequel, we further assume the following asynchronous model for the centralized batch solution \eqref{eqn:batchasyn}:
\begin{enumerate}
\item The random step-sizes $\{\bm{\mu}_k(i)\}$ satisfy the same properties as the asynchronous model for distributed diffusion networks described in Section \ref{I-sec:model} of Part I \cite{Zhao13TSPasync1}. In particular, the first and second-order moments of $\{\bm{\mu}_k(i)\}$ are constant and denoted by
\begin{align}
\label{eqn:randstepsizemeanentry}
\bar{\mu}_k & \defeq \E\,[\bm{\mu}_k(i)] \\
\label{eqn:randstepsizecoventry}
c_{\mu,k,\ell} & \defeq \E [(\bm{\mu}_k(i) - \bar{\mu}_k)(\bm{\mu}_\ell(i) - \bar{\mu}_\ell)]
\end{align}
for all $k$, $\ell$, and $i\ge0$, where the values of these moments are the same as those in \eqref{I-eqn:randstepsizemeanentry} and \eqref{I-eqn:randstepsizecoventry} from Part I \cite{Zhao13TSPasync1}.

\item The random fusion coefficients $\{\bm{\pi}_{k}(i)\}$ satisfy condition \eqref{eqn:convexcondition} at every iteration $i$. Moreover, the first and second-order moments of $\{\bm{\pi}_{k}(i)\}$ are denoted by
\begin{align}
\label{eqn:randcombinemeanentry}
\bar{\pi}_k & \defeq \E\,[\bm{\pi}_{k}(i)] \\
\label{eqn:randcombinecoventry}
c_{\pi,k,\ell} & \defeq \E [(\bm{\pi}_k(i) - \bar{\pi}_k)(\bm{\pi}_\ell(i) - \bar{\pi}_\ell)]
\end{align}
for all $k$, $\ell$, and $i\ge0$.

\item The random parameters $\{\bm{\mu}_k(i)\}$ and $\{\bm{\pi}_{k}(i)\}$ are mutually-independent and independent of any other random variable.
\end{enumerate}
We collect the fusion coefficients into the vector:
\be
\label{eqn:piidef}
\bm{\pi}_i \defeq \col\{ \bm{\pi}_1(i), \bm{\pi}_2(i), \dots, \bm{\pi}_N(i)\}
\ee
Then, condition \eqref{eqn:convexcondition} implies that $\bm{\pi}_i^\T \one_N = 1$. By \eqref{eqn:randcombinemeanentry} and \eqref{eqn:randcombinecoventry}, the mean and covariance matrix of $\bm{\pi}_i$ are given by
\begin{align}
\label{eqn:barpidef}
\bar{\pi} & \defeq \E\,(\bm{\pi}_i) = \col\{ \bar{\pi}_1, \bar{\pi}_2, \dots, \bar{\pi}_N \} \\
\label{eqn:Cpidef}
C_\pi & \defeq \E\,[(\bm{\pi}_i - \bar{\pi}) (\bm{\pi}_i - \bar{\pi})^\T] = \begin{bmatrix}
c_{\pi,1,1}  &   \dots  &  c_{\pi,1,N} \\
\vdots  &  \ddots  &  \vdots \\
c_{\pi,N,1}  &  \dots  &  c_{\pi,N,N} \\
\end{bmatrix}
\end{align}

\begin{lemma}[Properties of moments of $\{\bm{\pi}_k(i)\}$]
\label{lemma:conditionalmoments}
The first and second-order moments of $\{\bm{\pi}_k(i)\}$ defined in \eqref{eqn:randcombinemeanentry} and \eqref{eqn:randcombinecoventry} satisfy
\be 
\label{eqn:momentspikicondtions}
\sum_{k=1}^{N} \bar{\pi}_k = 1, \quad \bar{\pi}_k \ge 0, \quad \sum_{k=1}^{N} c_{\pi,k,\ell} = 0, \quad \sum_{\ell=1}^{N} c_{\pi,k,\ell} = 0
\ee
for any $k$ and $\ell$.
\end{lemma}
\begin{IEEEproof}
Using \eqref{eqn:barpidef} and \eqref{eqn:Cpidef} and the fact that $C_\pi$ is symmetric, conditions \eqref{eqn:momentspikicondtions} require
\be
\label{eqn:momentspikicondtionscompact}
\bar{\pi}^\T \one_N = 1, \qquad C_\pi \one_N = 0
\ee
The first equation in \eqref{eqn:momentspikicondtionscompact} is straightforward from \eqref{eqn:convexcondition}. The second condition in \eqref{eqn:momentspikicondtionscompact} is true because
\be
C_\pi \one_N = [\E(\bm{\pi}_i \bm{\pi}_i^\T | \bm{w}_{\c,i-1}) - \bar{\pi} \bar{\pi}^\T ] \one_N = 0
\ee
where we used the fact that $\bm{\pi}_i^\T \one_N = 1$ and $\bar{\pi}^\T \one_N = 1$.
\end{IEEEproof}

We next examine the stability and steady-state performance of the asynchronous batch algorithm \eqref{eqn:batchasyn}, and then compare its performance with that of the asynchronous distributed diffusion strategy.

\section{Performance of the Centralized Solution}
\label{sec:performance}
Following an argument similar to that given in Section \ref{I-sec:meansquarestability} of Part I \cite{Zhao13TSPasync1}, we can derive from \eqref{eqn:batchasyn1}--\eqref{eqn:batchasyn2} the following error recursion for the asynchronous centralized implementation:
\begin{subequations}
\begin{align}
\label{eqn:errorrecursionbatchasyn1}
\ubar{\wt{\bm{\psi}}}_{k,i} & = [ I_{2M} - \bm{\mu}_k(i) \bm{H}_{k,i-1} ]
\ubar{\wt{\bm{w}}}_{\c,i-1} + \ubar{\bm{s}}_{k,i} \\
\label{eqn:errorrecursionbatchasyn2}
\ubar{\wt{\bm{w}}}_{\c,i} & = \sum_{k=1}^{N} \bm{\pi}_{k}(i) \, \ubar{\wt{\bm{\psi}}}_{k,i}
\end{align}
\end{subequations}
where 
\begin{align}
\ubar{\wt{\bm{w}}}_{\c,i} & \defeq \ubar{\mbbT}(\wt{\bm{w}}_{\c,i}) \\
\ubar{\wt{\bm{\psi}}}_{k,i} & \defeq \ubar{\mbbT}(\wt{\bm{\psi}}_{k,i}) \\
\ubar{\bm{v}}_{k,i}({\bm{w}}_{\c,i-1}) & \defeq \ubar{\mbbT}(\bm{v}_{k,i}({\bm{w}}_{\c,i-1}))
\end{align}
and the mapping $\ubar{\mbbT}(\cdot)$ is from \eqref{I-eqn:ubardef} in Part I \cite{Zhao13TSPasync1}. Moreover,
\begin{align}
\label{eqn:Hkidef}
\bm{H}_{k,i-1} & \defeq \int_{0}^{1} \nabla_{\ubar{w}\ubar{w}^*}^2 J_k(\ubar{w}^o - t\ubar{\wt{\bm{w}}}_{\c,i-1})\,dt \\
\label{eqn:skidef}
\ubar{\bm{s}}_{k,i} & \defeq \bm{\mu}_k(i) \ubar{\bm{v}}_{k,i}({\bm{w}}_{\c,i-1})
\end{align}
We can merge \eqref{eqn:errorrecursionbatchasyn1} and \eqref{eqn:errorrecursionbatchasyn2} to find that the error dynamics of  \eqref{eqn:batchasyn} evolves according to the following recursion:
\be
\label{eqn:batcherrorrecursion}
\ubar{\wt{\bm{w}}}_{\c,i} = \left[ I_{2M} - \sum_{k=1}^{N} \bm{\pi}_{k}(i)\bm{\mu}_k(i) \bm{H}_{k,i-1} \right] \ubar{\wt{\bm{w}}}_{\c,i-1} + \ubar{\bm{s}}_{i}
\ee
where
\be
\label{eqn:sidef}
\ubar{\bm{s}}_i \defeq \sum_{k=1}^{N} \bm{\pi}_{k}(i) 
\ubar{\bm{s}}_{k,i} = \sum_{k=1}^{N} \bm{\pi}_{k}(i) \bm{\mu}_k(i) \ubar{\bm{v}}_{k,i}({\bm{w}}_{\c,i-1})
\ee

\subsection{Mean-Square and Mean-Fourth-Order Stability}

To maintain consistency with the notation used in Parts I \cite{Zhao13TSPasync1} and II \cite{Zhao13TSPasync2}, we shall employ the same auxiliary quantities in these parts for the centralized batch solution \eqref{eqn:batchasyn} with minor adjustments whenever necessary. For example, the error quantity $\wt{\bm{w}}_{k,i}$ used before in Parts I \cite{Zhao13TSPasync1} and II \cite{Zhao13TSPasync2} for the error vector at agent $k$ at time $i$ in the distributed implementation is now replaced by $\wt{\bm{w}}_{\c,i}$, with a subscript $c$, for the error vector of the centralized solution at time $i$. Thus, we let
\be
\label{eqn:epsilonidef}
\epsilon^2(i) \defeq \E\|\wt{\bm{w}}_{\c,i}\|^2 = \frac{1}{2}  \E\|\ubar{\wt{\bm{w}}}_{\c,i}\|^2
\ee
denote the MSD for the centralized solution $\wt{\bm{w}}_{\c,i}$.

\begin{theorem}[Mean-square stability]
\label{theorem:stability}
The mean-square stability of the asynchronous centralized implementation \eqref{eqn:batchasyn} reduces to studying the convergence of the recursive inequality:
\be
\label{eqn:meansquarerecursiveinequality}
\epsilon^2(i) \le \beta \cdot \epsilon^2(i-1) + \theta \sigma_v^2
\ee
where the parameters $\{\beta, \theta, \sigma_v^2\}$ are from \eqref{I-eqn:betadef}, \eqref{I-eqn:thetadef}, and \eqref{I-eqn:vkicond_var} in Part I \cite{Zhao13TSPasync1}, respectively.  The model \eqref{eqn:meansquarerecursiveinequality} is stable if condition
\be
\label{eqn:meansquarestablecondition}
\boxed{
\frac{\bar{\mu}_k^{(2)}}{\bar{\mu}_k^{(1)}} < \frac{\lambda_{k,\min}}{\lambda_{k,\max}^2 + \alpha}
}
\ee
holds for all $k$, where the parameters $\{ \lambda_{k,\min}, \lambda_{k, \max}, \alpha \}$ are from Assumptions \ref{I-asm:boundedHessian} and \ref{I-asm:gradientnoise} of Part I \cite{Zhao13TSPasync1}, respectively. When condition \eqref{eqn:meansquarestablecondition} holds, an upper bound on the steady-state MSD is given by
\be
\label{eqn:boundMSD}
\boxed{
\limsup_{i\rightarrow\infty} \E\,\|\wt{\bm{w}}_{\c,i}\|^2 \le b \cdot \nu
}
\ee
where $\nu$ is given by \eqref{eqn:nudef} and $b$ is a constant defined by \eqref{I-eqn:mumaxdef} from Part I \cite{Zhao13TSPasync1}.
\end{theorem}
\begin{IEEEproof}
Since the centralized solution \eqref{eqn:batchasyn}, or, equivalently, \eqref{eqn:batchasyn1}--\eqref{eqn:batchasyn2}, can  be viewed as a distributed solution over \emph{fully}-connected networks \cite{Zhao13ICASSP}, Theorem \ref{I-theorem:meansquarestability} from Part I \cite{Zhao13TSPasync1} can be applied directly. The result then follows from the fact that $\nu_o \le \nu$ by \eqref{I-eqn:nuandnuo} in Part I \cite{Zhao13TSPasync1}.
\end{IEEEproof}

Comparing the above result to Theorem 1 in Part I \cite{Zhao13TSPasync1}, we observe that the mean-square stability of the centralized solution \eqref{eqn:batchasyn} and the distributed asynchronous solution (39a)--(39b) from Part I \cite{Zhao13TSPasync1} is governed by the same model \eqref{eqn:meansquarerecursiveinequality}. Therefore, the same condition \eqref{eqn:meansquarestablecondition} guarantees the stability for both strategies and leads to the same MSD bound \eqref{eqn:boundMSD}.

\begin{theorem}[Stability of fourth-order error moment]
\label{theorem:4thmomentstability}
If
\be
\label{eqn:fourthorderstabilitycond}
\boxed{
\frac{\sqrt{ \bar{\mu}_k^{(4)} }}{ \bar{\mu}_k^{(1)} } < \frac{ \lambda_{k, \min} }{ 3\lambda_{k, \max}^2 + 4\alpha }
}
\ee
holds for all $k$, then the fourth-order moment of the error $\wt{\bm{w}}_{c,i}$ is asymptotically bounded by
\be
\label{eqn:bounded4thorderErrors}
\boxed{
\limsup_{i\rightarrow\infty} \E \| \wt{\bm{w}}_{c,i} \|^4 \le b_4^2 \cdot \nu^2
}
\ee
where the parameter $\nu$ is given by \eqref{eqn:nudef}, and $b_4$ is a constant defined by \eqref{I-eqn:nunewdef} of Part I \cite{Zhao13TSPasync1}.
\end{theorem}

\begin{IEEEproof}
This result follows from Theorem \ref{I-theorem:4thmoments} of Part I \cite{Zhao13TSPasync1} because the centralized solution \eqref{eqn:batchasyn}, or, equivalently, \eqref{eqn:batchasyn1}--\eqref{eqn:batchasyn2}, can be viewed as a distributed solution over \emph{fully}-connected networks \cite{Zhao13ICASSP}.
\end{IEEEproof}

An alternative method to investigate the stability conditions for the centralized solution \eqref{eqn:batchasyn} is to view it as a stochastic gradient descent iteration for a \emph{standalone} agent (i.e., a singleton network with $N = 1$) \cite{Sayed13Chapter, Sayed13SPM, Sayed14PROC}.

\subsection{Long Term Error Dynamics}
Using an argument similar to the one in Section \ref{II-subsec:longtermhighprob} from Part II \cite{Zhao13TSPasync2}, the original 
The original error recursion \eqref{eqn:batcherrorrecursion} can be rewritten as
\be
\label{eqn:batcherrorrecursion1}
\ubar{\wt{\bm{w}}}_{\c,i} = \left[ I_{2M} - \sum_{k=1}^{N} \bm{\pi}_{k}(i) \bm{\mu}_k(i) H_k \right] \ubar{\wt{\bm{w}}}_{\c,i-1} + \ubar{\bm{s}}_{i} + \bm{d}_i
\ee
where
\be
\bm{d}_i \defeq \sum_{k=1}^{N} \bm{\pi}_{k}(i) \bm{\mu}_k(i) (H_k - \bm{H}_{k,i-1}) \ubar{\wt{\bm{w}}}_{\c,i-1}
\ee
Then, under condition \eqref{eqn:fourthorderstabilitycond}, 
\be
\label{eqn:boundnormdi}
\limsup_{i\rightarrow\infty} \E\| \bm{d}_i \|^2 \le O(\nu^4)
\ee
where $\nu$ is given by \eqref{eqn:nudef}. 

\begin{assumption}[{Small step-sizes}]
\label{asm:smallstepsizes} 
The parameter $\nu$ from \eqref{eqn:nudef} is sufficiently small such that
\be
\nu < \min_k \frac{ \lambda_{k, \min} }{ 3\lambda_{k, \max}^2 + 4\alpha } < 1
\ee
\hfill \IEEEQED
\end{assumption}

Under Assumption \ref{asm:smallstepsizes}, condition \eqref{eqn:fourthorderstabilitycond} holds. Let
\begin{align}
\label{eqn:Bidef}
\bm{B}_i & \defeq \sum_{k=1}^{N} \bm{\pi}_{k}(i) \bm{D}_{k,i} \\
\label{eqn:Dkidef}
\bm{D}_{k,i} & \defeq I_{2M} - \bm{\mu}_k(i) H_k
\end{align}
where $\bm{B}_i$ is Hermitian positive semi-definite. Since we are interested in examining the asymptotic performance of the asynchronous batch solution, we can again call upon the same argument from Section \ref{II-subsec:longtermhighprob} of Part II \cite{Zhao13TSPasync2} and use result \eqref{eqn:boundnormdi} to conclude that we can assess the performance of \eqref{eqn:batcherrorrecursion1} by working with the following \emph{long-term} model, which holds for large enough $i$:
\be
\label{eqn:batcherrorrecursionapprox}
\ubar{\wt{\bm{w}}}_{\c,i}' = \bm{B}_i \cdot \ubar{\wt{\bm{w}}}_{\c,i-1}' + \ubar{\bm{s}}_i, \;\; i \gg 1 
\ee
In model \eqref{eqn:batcherrorrecursionapprox}, we ignored the $O(\nu^2)$ term $\bm{d}_i$ according to \eqref{eqn:boundnormdi}, and we are using $\bm{w}_{c,i}'$ to denote the estimate obtained from this long-term model. Note that the driving noise term $\ubar{\bm{s}}_i$ in \eqref{eqn:batcherrorrecursionapprox} is extraneous and imported from the original error recursion \eqref{eqn:batcherrorrecursion}.

\begin{theorem}[Bounded mean-square gap]
\label{theorem:gapmeansquare}
Under Assumption \ref{asm:smallstepsizes}, the mean-square gap from the original error recursion \eqref{eqn:batcherrorrecursion} to the long-term model \eqref{eqn:batcherrorrecursionapprox} is asymptotically bounded by 
\be
\limsup_{i \rightarrow \infty} \E \|\ubar{\wt{\bm{w}}}_{c,i} - \ubar{\wt{\bm{w}}}_{c,i}' \|^2 \le O(\nu^2)
\ee
where $\nu$ is given by \eqref{eqn:nudef}.
\end{theorem}
\begin{IEEEproof}
This result follows from Theorem \ref{II-theorem:gapmeansquare} of Part II \cite{Zhao13TSPasync2} since the centralized solution \eqref{eqn:batchasyn} can be viewed as a distributed solution over \emph{fully}-connected networks \cite{Zhao13ICASSP}.
\end{IEEEproof}

\subsection{Mean Error Recursion}
By taking the expectation of both sides of \eqref{eqn:batcherrorrecursionapprox}, and using the fact that $\E(\ubar{\bm{s}}_i) = 0$, we conclude that the mean error satisfies the recursion: 
\be
\label{eqn:meanerrorrecursion}
\E\,\ubar{\wt{\bm{w}}}_{\c,i}' = \bar{B} \cdot \E\,\ubar{\wt{\bm{w}}}_{\c,i-1}'
\ee
for large enough $i$, where
\begin{align}
\label{eqn:Bmeandef}
\bar{B} & \defeq \E (\bm{B}_i) =  \sum_{k=1}^{N} \bar{\pi}_k \bar{D}_k \\
\label{eqn:Dkmeandef}
\bar{D}_k & \defeq \E(\bm{D}_{k,i}) = I_{2M} - \bar{\mu}_k H_k
\end{align}
The convergence of recursion \eqref{eqn:meanerrorrecursion} requires the stability of $\bar{B}$. It is easy to verify that $\{\bar{B}, \bar{D}_k\}$ are Hermitian. Using \eqref{eqn:momentspikicondtions} and Jensen's inequality, we get from \eqref{eqn:Bmeandef} that $\rho(\bar{B}) \le \max_k \rho(\bar{D}_k)$. As we showed in \eqref{II-eqn:meanstabecondition} from Part II \cite{Zhao13TSPasync1}, if condition \eqref{eqn:meansquarestablecondition} holds, then $\rho(\bar{D}_k) < 1$ for all $k$. Therefore, it follows from Assumption \ref{asm:smallstepsizes} that
\be
\label{eqn:limmeanerror}
\lim_{i\rightarrow\infty} \E\,\ubar{\wt{\bm{w}}}_{\c,i}' = 0
\ee
which implies that the long-term model \eqref{eqn:batcherrorrecursionapprox} is the asymptotically centered version of the original error recursion \eqref{eqn:batcherrorrecursion}.

\subsection{Error Covariance Recursion}
Let $\F_{i-1}$ denote the filtration that represents all information available up to iteration $i-1$. Then we deduce from \eqref{eqn:batcherrorrecursionapprox} that for large enough $i$:
\begin{align}
\label{eqn:errorcovrecursion2}
\E(\ubar{\wt{\bm{w}}}_{\c,i}' \ubar{\wt{\bm{w}}}_{\c,i}'^* | \F_{i-1} ) & = \E(\bm{B}_i \ubar{\wt{\bm{w}}}_{\c,i-1}' 
\ubar{\wt{\bm{w}}}_{\c,i-1}'^* \bm{B}_i | \F_{i-1} ) 
 + \E(\ubar{\bm{s}}_i\ubar{\bm{s}}_i^* | \F_{i-1} )
\end{align}
where the cross terms that involve $\ubar{\bm{s}}_i$ disappear because $\E(\ubar{\bm{s}}_i^* \bm{B}_i \ubar{\wt{\bm{w}}}_{\c,i-1}' | \F_{i-1} ) = 0$ by the gradient noise model from Assumption \ref{II-asm:gradientnoisestronger} of Part II \cite{Zhao13TSPasync2}. Vectorizing both sides of \eqref{eqn:errorcovrecursion2} and taking expectation, we obtain
\be
\label{eqn:errorcovrecursion3}
\E\,[ (\ubar{\wt{\bm{w}}}_{\c,i}'^*)^\T \kron 
\ubar{\wt{\bm{w}}}_{\c,i}'  ] = F_\c \cdot \E\,[ ( \ubar{\wt{\bm{w}}}_{\c,i-1}'^* )^\T \kron \ubar{\wt{\bm{w}}}_{\c,i-1}' ] + y_{\c,i}
\ee
where
\begin{align}
\label{eqn:Fcdef}
F_\c & \defeq  \E\,[ \bm{B}_i^\T \kron \bm{B}_i ] \\
\label{eqn:ycidef}
y_{\c,i} & \defeq \E[ (\ubar{\bm{s}}_i^*)^\T \kron \ubar{\bm{s}}_i ] 
\end{align}
Let further
\be
\label{eqn:Hcdef}
H_\c \defeq \sum_{k=1}^{N} \bar{\pi}_k \bar{\mu}_k H_k = O(\nu)
\ee
where $\{ H_k \}$ are from \eqref{II-eqn:Hkidef} of Part II \cite{Zhao13TSPasync2}.

\begin{lemma}[Properties of $F_\c$]
\label{lemma:propertiesFc}
The matrix $F_\c$ defined by \eqref{eqn:Fcdef} is Hermitian and can be expressed as
\be
\label{eqn:Fcexpression}
F_\c = \sum_{\ell = 1}^{N} \sum_{k=1}^{N} (\bar{\pi}_\ell \bar{\pi}_k  +  c_{\pi,\ell,k})(\bar{D}_\ell^\T \kron \bar{D}_k  +  c_{\mu,\ell,k} H_\ell^\T \kron H_k)
\ee
If condition \eqref{eqn:meansquarestablecondition} holds, then $F_\c$ is stable and
\be
\label{eqn:rhoFcexpression}
\rho(F_\c) = [1-\lambda_{\min}(H_\c)]^2 + O(\nu^2)
\ee
where $H_\c$ is given by \eqref{eqn:Hcdef}, and $[1-\lambda_{\min}(H_\c)]^2 = 1- O(\nu)$ under Assumption \ref{asm:smallstepsizes}. Moreover,
\be
\label{eqn:IminusFcinvOrder}
\| (I_{4M^2} - F_\c)^{-1} \| = O(\nu^{-1})
\ee 
\end{lemma}
\begin{IEEEproof}
See Appendix \ref{app:propertiesFc}.
\end{IEEEproof}

\begin{theorem}[Error covariance recursion]
\label{theorem:errorcovariance}
For sufficiently large $i$, the vectorized error covariance for the long-term model \eqref{eqn:batcherrorrecursionapprox} satisfies the following relation:
\be
\label{eqn:errorcovrecursion}
z_{\c,i} = F_\c \cdot z_{\c,i-1} + y_{\c,i}, \;\; i \gg 1
\ee
where $F_\c$ and $y_{\c,i}$ are from \eqref{eqn:Fcdef} and \eqref{eqn:ycidef}, respectively, and
\be
\label{eqn:zcidef}
z_{\c,i} \defeq \E\,[(\ubar{\wt{\bm{w}}}_{\c,i}'^*)^\T \kron \ubar{\wt{\bm{w}}}_{\c,i}' ]
\ee
Recursion \eqref{eqn:errorcovrecursion} is convergent if condition \eqref{eqn:meansquarestablecondition} holds, and its convergence rate is dominated by $[1-\lambda_{\min}(H_\c)]^2 = 1 - O(\nu)$ under Assumption \ref{asm:smallstepsizes}.
\end{theorem}

\begin{IEEEproof}
Equation \eqref{eqn:errorcovrecursion} follows from \eqref{eqn:errorcovrecursion3}. Recursion \eqref{eqn:errorcovrecursion} converges if, and only if, the matrix $F_\c$ is stable. By Lemma \ref{lemma:propertiesFc}, we know that $\rho(F_\c) < 1$ if condition \eqref{eqn:meansquarestablecondition} holds and, moreover, the convergence rate of recursion \eqref{eqn:errorcovrecursion} is determined by $\rho(F_\c) = [1 - \lambda_{\min}(H_\c)]^2 + O(\nu^2)$.
\end{IEEEproof}

\subsection{Steady-State MSD}
At steady-state as $i \rightarrow \infty$, we get from \eqref{eqn:IminusFcinvOrder} and \eqref{eqn:errorcovrecursion} that
\begin{align}
\label{eqn:zcinfdef}
z_{\c,\infty} & \defeq \vecm\left( \lim_{i\rightarrow\infty} \E \ubar{\wt{\bm{w}}}_{\c,i}' \ubar{\wt{\bm{w}}}_{\c,i}'^* \right) \nn \\
& = (I_{4M^2}  -  F_\c)^{-1} \cdot \lim_{i\rightarrow\infty} y_{\c, i} 
\end{align}
Using $z_{\c,\infty}$, we can determine the value of any steady-state weighted mean-square-error metric for the long-term model \eqref{eqn:batcherrorrecursionapprox} as follows:
\begin{align}
\label{eqn:weightedMSEmetric}
\lim_{i\rightarrow\infty} \E\|\wt{\bm{w}}_{\c,i}'\|_\Sigma^2 & = \frac{1}{2} \lim_{i\rightarrow\infty} \Tr[ \E(\ubar{\wt{\bm{w}}}_{\c,i}' \ubar{\wt{\bm{w}}}_{\c,i}'^* ) \Sigma] \nn \\
& = \frac{1}{2} z_{\c,\infty}^* \vecm(\Sigma)
\end{align}
where we used the fact that $\Tr(AB) = [\vecm(A^*)]^* \vecm(B)$, and $\Sigma$ is an arbitrary Hermitian positive semi-definite weighting matrix. The steady-state MSD for the original error recursion \eqref{eqn:batcherrorrecursion1} is defined by
\be
\label{eqn:MSDbatchdef}
\MSD^{\cent} \defeq \lim_{i\rightarrow\infty} \E\|\wt{\bm{w}}_{\c,i}\|^2 = \lim_{i\rightarrow\infty} \frac{1}{2}\E\|\ubar{\wt{\bm{w}}}_{\c,i}\|^2
\ee
Therefore, by setting $\Sigma = I_{2M}$ in \eqref{eqn:weightedMSEmetric} and using Theorem \ref{theorem:gapmeansquare}, it is easy to verify by following an argument similar to the proof of Theorem \ref{II-theorem:steadystateMSD} from Part II \cite{Zhao13TSPasync2} that
\be
\label{eqn:gapMSD}
\MSD^{\cent} = \lim_{i\rightarrow\infty} \E\|\wt{\bm{w}}_{\c,i}'\|^2 + O(\nu^{3/2})
\ee
Introduce
\be
\label{eqn:Rcdef}
R_\c \defeq \sum_{k=1}^{N} (\bar{\pi}_k^2 + c_{\pi,k,k} ) (\bar{\mu}_k^2 + c_{\mu,k,k} ) R_k = O(\nu^2)
\ee
where $\{ R_k \}$ are from \eqref{II-eqn:Rkdef} of Part II \cite{Zhao13TSPasync2}. Then, using \eqref{eqn:weightedMSEmetric} and \eqref{eqn:gapMSD}, we arrive the following result.

\begin{theorem}[Steady-state MSD]
\label{theorem:MSD}
The steady-state MSD for the asynchronous centralized (batch) solution \eqref{eqn:batchasyn} is given by
\be
\label{eqn:MSDbatchexpression}
\MSD^{\cent} = \frac{1}{2} [ \vecm( R_\c ) ]^* (I_{4M^2} - F_\c)^{-1} \vecm(I_{2M}) + O(\nu^{1+\gamma_o})
\ee
where $0 < \gamma_o \le 1/2$ is from \eqref{II-eqn:gammaodef} of Part II \cite{Zhao13TSPasync2}. Expression \eqref{eqn:MSDbatchexpression} can be further reworked to yield
\be
\label{eqn:MSDbatchapprox}
\MSD^{\cent} = \frac{1}{4} \Tr(H_\c^{-1} R_\c) + O(\nu^{1 + \gamma_o})
\ee
where the first term on the RHS is in the order of $\nu$ and therefore dominates the $O(\nu^{1 + \gamma_o})$ term under Assumption \ref{asm:smallstepsizes}.
\end{theorem}
\begin{IEEEproof}
See Appendix \ref{app:MSD}.
\end{IEEEproof}

\subsection{Results for the Synchronous Centralized Solution}

We may also consider a synchronous centralized (batch) implementation for solving the same problem \eqref{eqn:globalcost}. It would take the following form:
\be
\label{eqn:batchsyn}
\bm{w}_{\c,i} = \bm{w}_{\c,i-1} - \sum_{k=1}^{N} \pi_k \mu_k \wh{\nabla_{w^*}J_k}(\bm{w}_{\c,i-1})
\ee
where the $\{\mu_k\}$ are now deterministic nonnegative step-sizes and the $\{\pi_k\}$ are nonnegative fusion coefficients that satisfy $\sum_{k=1}^{N} \pi_k = 1$. The synchronous batch solution can be viewed as a special case of the asynchronous batch solution \eqref{eqn:batchasyn} when the random step-sizes and fusion coefficients assume constant values. If the covariances $\{ c_{\mu,k,k} \}$ and $\{ c_{\pi, k, k} \}$ are set to zero, then the asynchronous solution \eqref{eqn:batchasyn} will reduce into a synchronous solution that employs the constant parameters $\{ \bar{\mu}_k \}$ and $\{ \bar{\pi}_k \}$. The previous stability and performance results can be specialized to the synchronous batch implementation under these conditions.

It is easy to verify that the mean error recursion for the synchronous solution with parameters $\{ \bar{\mu}_k \}$ and $\{ \bar{\pi}_k \}$ is identical to \eqref{eqn:meanerrorrecursion}. The mean convergence rate for the long-term model is still determined by $\rho(\bar{B})$, where $\bar{B}$ is given by \eqref{eqn:Bmeandef}. The mean square convergence rate for the long-term model is determined by $\rho(F_\c')$ where
\be
\label{eqn:Fcprimedef}
F_\c' \defeq \sum_{k=1}^{N} \sum_{\ell=1}^{N} \bar{\pi}_\ell \bar{\pi}_k (\bar{D}_\ell^\T \kron \bar{D}_k)
\ee
It follows that
\be
\rho(F_\c') = [1 - \lambda_{\min}(H_c)]^2 = 1 - O(\nu)
\ee
The steady-state MSD is given by
\be
\label{eqn:MSDsyncbatchapprox}
\MSD_{\sync}^{\cent} = \frac{1}{4} \Tr(H_\c^{-1} R_\c') + O(\nu^{1 + \gamma_o})
\ee
where
\be
R_\c' \defeq \sum_{k=1}^{N} \bar{\pi}_k^2 \bar{\mu}_k^2 R_k, \quad \| R_\c' \| = O(\nu^2)
\ee
and $\Tr(H_\c^{-1} R_\c') = O(\nu)$.

\section{Comparison I: Distributed vs. Centralized Strategies}

In this section, we compare the mean-square performance of the distributed diffusion strategy \eqref{I-eqn:atcasync1}--\eqref{I-eqn:atcasync2} from Part I \cite{Zhao13TSPasync1}, namely,
\begin{subequations}
\begin{align}
\label{eqn:atcasync1}
{\bm{\psi}}_{k,i} & = {\bm{w}}_{k,i-1} - \bm{\mu}_k(i) \wh{\nabla_{w^*}J_k}(\bm{w}_{k,i-1}) \\
\label{eqn:atcasync2}
{\bm{w}}_{k,i} & = \sum_{\ell\in\bm{\Ncal}_{k,i}} \bm{a}_{\ell k}(i)\,{\bm{\psi}}_{\ell,i}
\end{align}
\end{subequations}
with the centralized (batch) solution described by \eqref{eqn:batchasyn}. We establish the important conclusion that if the combination matrix is primitive (Assumption \ref{II-asm:connected} in Part II \cite{Zhao13TSPasync2}), then the asynchronous network is able to achieve almost the same mean-square performance as the centralized (batch) solution for sufficiently small step-sizes. In other words, diffusion strategies are \emph{efficient} mechanisms to perform continuous adaptation and learning tasks over networks even in the presence of various sources of random failures.

\subsection{Adjusting Relevant Parameters}
First, however, we need to describe the conditions that are necessary for a \emph{fair} and meaningful comparison between the distributed and centralized implementations. This is because the two implementations use different parameters. Recall that the agents in the distributed network \eqref{eqn:atcasync1}--\eqref{eqn:atcasync2} employ random combination coefficients $\{\bm{a}_{\ell k}(i)\}$ to aggregate information from neighborhoods using random step-sizes $\{\bm{\mu}_k(i)\}$. The random parameters $\{\bm{a}_{\ell k}(i), \bm{\mu}_k(i)\}$ are assumed to satisfy the model described in Section \ref{I-sec:model} from Part I \cite{Zhao13TSPasync1}. On the other hand, the centralized batch solution \eqref{eqn:batchasyn} uses random combination coefficients $\{\bm{\pi}_k(i)\}$ to fuse the information from all agents in the network, and then performs updates using random step-sizes $\{\bm{\mu}_k(i)\}$. The random parameters $\{\bm{\pi}_k(i), \bm{\mu}_k(i)\}$ are assumed to satisfy the conditions specified in Section \ref{subsec:model} of this part. In general, the two sets of random parameters, i.e., $\{\bm{a}_{\ell k}(i), \bm{\mu}_k(i)\}$ for distributed strategies and $\{\bm{\pi}_k(i), \bm{\mu}_k(i)\}$ for centralized strategies, are not necessarily related. Therefore, in order to make a meaningful comparison between the distributed and centralized strategies, we need to introduce connections between these two sets of parameters. This is possible because the parameters play similar roles.

From the previous analysis in Section \ref{II-sec:lowrank} of Part II \cite{Zhao13TSPasync2}, we know that the first and second-order moments of $\{\bm{a}_{\ell k}(i), \bm{\mu}_k(i)\}$ determine the mean-square performance of diffusion networks. Likewise, from the analysis in Section \ref{sec:performance} of this part, we know that the first and second-order moments of $\{\bm{\pi}_k(i), \bm{\mu}_k(i)\}$ determine the mean-square performance of centralized solutions. Therefore, it is sufficient to introduce connections between the first and second-order moments of these random parameters. For the random step-size parameters, we assumed in \eqref{I-eqn:randstepsizemeanentry} and \eqref{I-eqn:randstepsizecoventry} from Part I \cite{Zhao13TSPasync1} and in \eqref{eqn:randstepsizemeanentry} and \eqref{eqn:randstepsizecoventry} from this part that their first and second-order moments are \emph{constant} and that their values coincide with each other, i.e., $\bar{\mu}_k$ from \eqref{I-eqn:randstepsizemeanentry} in Part I \cite{Zhao13TSPasync1} coincides with $\bar{\mu}_k$ from \eqref{eqn:randstepsizemeanentry} in this part, and similarly for $c_{\mu,k,\ell}$. This requirement is obviously reasonable.

The connection that we need to enforce between the moments of the combination coefficients $\{\bm{a}_{\ell k}(i)\}$ and $\{\bm{\pi}_k(i)\}$, while reasonable again, is less straightforward to explain. This is because the $\{\bm{a}_{\ell k}(i)\}$ form a random matrix $\bm{A}_i = [\bm{a}_{\ell k}(i)]_{k,\ell=1}^{N}$ of size $N\times N$, while the $\{\bm{\pi}_k(i)\}$ only form a random vector $\bm{\pi}_i = [\bm{\pi}_k(i)]_{k=1}^{N}$ of size $N\times1$. From the result of Corollary \ref{II-corollary:steadystateMSD} in Part II \cite{Zhao13TSPasync2} though, we know that the mean-square performance of the \emph{primitive} diffusion network does not \emph{directly} depend on the moments of $\bm{A}_i$, namely, its mean $\bar{A}$ and its Kronecker covariance $C_A$; instead, the performance depends on the Perron eigenvector (the unique right eigenvector corresponding to the eigenvalue at one for primitive left-stochastic matrices \cite{BermanPF, Pillai05SPM}). If, for example, we compare expression \eqref{II-eqn:steadyMSDapprox} from Part II \cite{Zhao13TSPasync2} for asynchronous networks with expression \eqref{eqn:Fcexpression} from this part, we conclude that it is sufficient to relate the vectors $\{\bar{p}, p\}$ defined in \eqref{II-eqn:pdef} and \eqref{II-eqn:barpdef} from Part II \cite{Zhao13TSPasync2} to the moments $\{\bar{\pi}_k, c_{\pi,k,\ell}\}$. Since $\bar{p}$ is the Perron eigenvector of the mean matrix $\bar{A}$, and the $\{\bar{\pi}_k\}$ are the means of $\{\bm{\pi}_k(i)\}$, we connect them by requiring
\be
\label{eqn:barpkandbarpik}
\bar{\pi}_k \equiv \bar{p}_k
\ee
for all $k$, where the $\{\bar{p}_k\}$ are the elements of $\bar{p}$. Likewise, since $p$ is the Perron eigenvector of the matrix $\bar{A} \kron \bar{A} + C_A = \E\,(\bm{A}_i \kron \bm{A}_i)$, which consists of the second-order moments, and $\{\bar{\pi}_k\bar{\pi}_\ell + c_{\pi,k,\ell} = \E\,[\bm{\pi}_k(i) \bm{\pi}_\ell(i)]\}$ are also the second-order moments, we connect them by requiring
\be
\label{eqn:piandp}
\bar{\pi}_k \bar{\pi}_\ell + c_{\pi,k,\ell} \equiv p_{k,\ell}
\ee
for all $k$ and $\ell$, where the $\{p_{k,\ell}\}$ are the elements of $p$ defined after \eqref{II-eqn:Ppdef} in Part II \cite{Zhao13TSPasync2}. When conditions \eqref{eqn:barpkandbarpik} and \eqref{eqn:piandp} are satisfied, then the mean-square convergence rates and steady-state MSD for the distributed and centralized solutions become identical. We establish this result in the sequel. Using \eqref{eqn:barpidef} and \eqref{eqn:Cpidef}, conditions \eqref{eqn:barpkandbarpik} and \eqref{eqn:piandp} can be rewritten as
\be
\label{eqn:barpiandbarpandCpiandPp}
\bar{\pi} \equiv \bar{p}, \qquad 
C_\pi + \bar{\pi}\bar{\pi}^\T \equiv P_p
\ee
where 
\be
P_p = \begin{bmatrix}
p_{1,1} & \dots & p_{1,N} \\
\vdots & \ddots & \vdots \\
p_{N,1} & \dots & p_{N,N} \\
\end{bmatrix}
\ee
is the symmetric matrix defined by \eqref{II-eqn:Ppdef} of Part II \cite{Zhao13TSPasync2}. It is worth noting that, since the Perron eigenvectors $p = \vecm(P_p)$ and $\bar{p}$ consist of positive entries, the corresponding quantities $\bar{\pi}$ and $C_\pi + \bar{\pi}\bar{\pi}^\T$ must also consist of positive entries --- we shall refer to the centralized solutions that satisfy this condition as \emph{primitive} centralized solutions. Clearly, the second requirement in \eqref{eqn:barpiandbarpandCpiandPp} is meaningful only if the difference $P_p - \bar{p}\bar{p}^\T$ results in a symmetric positive semi-definite matrix (and, hence, a covariance matrix) that also satisfies $C_{\pi}\one_N = 0$.

\subsection{Constructing Primitive Batch Solutions}
\label{subsec:construct}
Before comparing the performance of the centralized and distributed solutions under \eqref{eqn:barpiandbarpandCpiandPp}, we first answer the following important inquiry. Given a distributed primitive network with parameters $\{\bar{p}, P_p\}$, is it possible to determine a batch solution with parameters $\{\bar{\pi}, C_{\pi}\}$ satisfying \eqref{eqn:barpiandbarpandCpiandPp} such that the resulting $C_{\pi}$ is a symmetric and positive semi-definite matrix (and, therefore, has the interpretation of a valid covariance matrix)? The answer is in the affirmative as we proceed to explain. The following are auxiliary results in this direction.

\begin{lemma}[Positive semi-definite property]
\label{lemma:nonnegativity}
The matrix difference $P_p - \bar{p}\bar{p}^\T$ is symmetric positive semi-definite and satisfies $(P_p - \bar{p}\bar{p}^\T)\one_N = 0$ for any $\bar{p}$ and $P_p$ defined by \eqref{II-eqn:barpdef} and \eqref{II-eqn:Ppdef} from Part II \cite{Zhao13TSPasync2}.
\end{lemma}
\begin{IEEEproof}
See Appendix \ref{app:nonnegativity}.
\end{IEEEproof}

Therefore, starting from an asynchronous diffusion network with parameters $\{\bar{p}, P_p\}$, there exists an asynchronous batch solution with valid parameters $\{\bar{\pi}, C_{\pi}\}$ that satisfy \eqref{eqn:barpiandbarpandCpiandPp}. We now explain one way by which a random variable $\bm{\pi}_i$ can be constructed with the pre-specified moments $\{\bar{\pi}, C_{\pi}\}$. We first observe that in view of condition \eqref{eqn:piidef}, the random variable $\bm{\pi}_i$ is actually defined on the probability simplex in $\mbbR^{N\times1}$ \cite[p.~33]{Boyd04}:
\be
\label{eqn:simplexdef}
\Delta_N \defeq \{ x \in \mbbR^{N\times1}; x^\T\one_N = 1, x_k \ge 0, k = 1,\dots,N\}
\ee
If the moments $\{\bar{\pi}, C_{\pi}\}$ obtained from \eqref{eqn:barpiandbarpandCpiandPp} satisfy certain conditions, then there are several models in the literature that can be used to generate random vectors $\{\bm{\pi}_i\}$ according to these moments such as using the Dirichlet distribution \cite{Kotz00}, the Generalized Dirichlet distribution \cite{Connor69JASA, Aitchison85JRSS, Barndorff91JMA, Wong98AMC, Hankin10JSS, Chang10CCM, Wong10CSDA, Favaro11JSPI}, the Logistic-Normal distribution \cite{Aitchison80Bio, Aitchison85JRSS, Lenk88JASA}, or the Generalized inverse Gaussian distribution \cite{Barndorff91JMA, Seshadri92SPL}. Unfortunately, if the conditions for these models are not satisfied, no \emph{closed-form} probabilistic model is available for us to generate random variables on the probability simplex with pre-specified means and covariance matrices.

Nevertheless, inspired by the Markov Chain Monte Carlo (MCMC) method \cite{Andrieu03ML}, we describe one procedure to construct random variables \emph{indirectly} so that they are able to meet the desired moment requirements. In a manner similar to the argument used in Appendix \ref{app:nonnegativity}, we introduce a series of fictitious random combination matrices $\{\bm{A}_j';j\ge1\}$ that satisfy the asynchronous model introduced in Part I \cite{Zhao13TSPasync1}. We assume that the $\{\bm{A}_j';j\ge1\}$ are independently, identically distributed (i.i.d.) random matrices, and they are independent of any other random variable. Then, the mean and Kronecker-covariance matrices of $\bm{A}_j'$ for any $j$ are given by $\bar{A}$ and $C_A$, respectively. We further introduce the random matrix
\be
\label{eqn:Phiidef}
\bm{\Phi}_{i,t} \defeq \prod_{j=1}^t \bm{A}_j'
\ee
Similar to \eqref{eqn:PhikronPhi} and \eqref{eqn:Phi}, we can verify that
\be
\label{eqn:Phiinfmeanandvar}
\lim_{t\rightarrow\infty} \E(\bm{\Phi}_{i,t} ) = \bar{p} \one_{N}^\T, \quad 
\lim_{t\rightarrow\infty} \E(\bm{\Phi}_{i,t} \kron \bm{\Phi}_{i,t} ) = p \one_{N^2}^\T
\ee
Let
\be
\label{eqn:phidef}
\bm{\phi}_i \defeq \frac{1}{N} \left( \lim_{t\rightarrow\infty} \bm{\Phi}_{i,t}\right) \one_N
\ee
Then, the entries of $\bm{\phi}_i$ are nonnegative since the entries of $\bm{\Phi}_{i,t}$ are nonnegative. Using \eqref{eqn:Phiidef} and \eqref{eqn:phidef}, we have
\be
\one_N^\T \bm{\phi}_i = \frac{1}{N} \lim_{t\rightarrow\infty} \one_N^\T \left( \prod_{j=1}^t \bm{A}_j' \right) \one_N = 1
\ee
since each $\bm{A}_j'$ is left-stochastic. Therefore, $\bm{\phi}_i$ is a random variable defined on the probability simplex $\Delta_N$. By using \eqref{eqn:Phiinfmeanandvar} and the fact that $\one_{N} \kron \one_{N} = \one_{N^2}$, we have
\begin{align}
\E(\bm{\phi}_i) & = \frac{1}{N} (\bar{p} \cdot \one_{N}^\T) \one_N = \bar{p} \\
\E(\bm{\phi}_i \kron \bm{\phi}_i) & = \frac{1}{N^2} (p \cdot \one_{N^2}^\T) \one_{N^2} = p
\end{align}
Therefore,
\be
\E(\bm{\phi}_i) = \bar{p}, \qquad \Cov(\bm{\phi}_i) = P_p - \bar{p}\bar{p}^\T
\ee
where $P_p = \unvecm(p)$. In this way, we have been able to construct a random variable $\bm{\phi}_i$ whose support is the probability simplex $\Delta_N$ and whose mean vector and covariance matrix match the specification. The random variable $\bm{\phi}_i$ can then be used by the asynchronous centralized solution at time $i$, which would then enable a meaningful comparison with the asynchronous distributed solution.

Although unnecessary for our development, it is instructive to pose the converse question: Given a \emph{primitive} batch solution with parameters $\{\bar{\pi}, C_{\pi}\}$, is it always possible to determine a distributed solution with parameters $\{\bar{p}, P_{p}\}$ satisfying \eqref{eqn:barpiandbarpandCpiandPp} such that these parameters have the properties of Perron eigenvectors? In other words, given a primitive centralized solution, is it possible to determine a distributed solution on a \emph{partially}-connected network (otherwise the problem is trivial since fully-connected networks are equivalent to centralized solutions \cite{Zhao13ICASSP}) with equivalent performance levels? The answer to this question remains open. The challenge stems from the fact mentioned earlier that, in general, there is no systematic solution to generate distributions on the probability simplex with pre-specified first and second-order moments. The method of moments \cite{Gelman95JCGS}, which is an iterative solution, does not generally guarantee convergence and therefore, cannot ensure that a satisfactory distribution can be generated eventually.

\subsection{Comparing Performance}

From the mean error recursion in \eqref{II-eqn:meanerrorrecursion} of Part II \cite{Zhao13TSPasync2}, the mean convergence rate for the long-term model of the distributed diffusion strategy is determined by $\rho(\bar{\Bcal})$, where $\bar{\Bcal}$ is defined by \eqref{II-eqn:bigBmeandef} of Part II \cite{Zhao13TSPasync2}. From the mean error recursion \eqref{eqn:meanerrorrecursion} in this part, the mean convergence rate for the long-term model of the centralized batch solution is determined by $\rho(\bar{B})$, where $\bar{B}$ is given by \eqref{eqn:Bmeandef}. 

\begin{lemma}[Matching mean convergence rates]
\label{lemma:matchingmeanconvergencerate}
The mean convergence rates for the asynchronous distributed strategy and the centralized batch solution are almost the same. Specifically, it holds that
\be
\label{eqn:rhobigBandrhoBmean}
| \rho(\bar{\Bcal}) - \rho(\bar{B}) | \le O(\nu^{1+1/N})
\ee
where $\rho(\bar{\Bcal})$ and $\rho(\bar{B})$ are of the order of $1 - O(\nu)$.
\end{lemma}
\begin{IEEEproof}
See Appendix \ref{app:matchingmeanconvergencerate}.
\end{IEEEproof}

Likewise, from Theorem \ref{II-theorem:errorcovariancerecursion} of Part II \cite{Zhao13TSPasync2}, the mean-square convergence rate of the distributed diffusion strategy for large enough $i$ is determined by $\rho(\Fcal)$, where $\Fcal$ is from \eqref{II-eqn:bigFdef} of Part II \cite{Zhao13TSPasync2}. From Theorem \ref{theorem:errorcovariance} of this part, the mean-square convergence rate of the centralized (batch) solution is determined by $\rho(F_\c)$, where $F_\c$ is from \eqref{eqn:Fcdef}.

\begin{lemma}[Matching mean-square convergence rates]
\label{lemma:matchingmeansquareconvergencerate}
The mean-square convergence rates for the asynchronous distributed strategy and the centralized batch solution are almost the same. Specifically, it holds that
\be
\label{eqn:rhobigFandrhoFc} 
| \rho(\Fcal) - \rho(F_\c) | \le O(\nu^{1+1/N^2})
\ee
where $\rho(\Fcal)$ and $\rho(F_\c)$ are of the order of $1 - O(\nu)$.
\end{lemma}
\begin{IEEEproof}
From \eqref{eqn:Fcdef} and \eqref{eqn:piandp}, it is easy to verify that $F_\c = F$, where $F$ is from \eqref{II-eqn:Fdef} of Part II \cite{Zhao13TSPasync2}. Using Lemmas \ref{II-lemma:spectralF} and \ref{II-lemma:lowrank} from Part II \cite{Zhao13TSPasync2} then completes the proof. 
\end{IEEEproof}

The steady-state network MSD for the distributed diffusion strategy is given by \eqref{II-eqn:steadyMSDapprox} of Part II \cite{Zhao13TSPasync2}:
\be
\label{eqn:MSDdiffusion}
\MSD^{\dist} = \frac{1}{4}\Tr(H^{-1}R) + O(\nu^{1 + \gamma_o})
\ee
for some $0 < \gamma_o \le 1/2$ given by \eqref{II-eqn:gammaodef} of Part II \cite{Zhao13TSPasync2}. The steady-state MSD for the centralized batch solution is given by \eqref{eqn:MSDbatchapprox}.

\begin{lemma}[Matching MSD performance]
\label{lemma:matchingMSD}
At steady-state, the network MSD for the asynchronous distributed strategy and the MSD for the centralized batch solution are close to each other. Specifically, we have
\be
\label{eqn:MSDnetworkandMSDcent}
| \MSD^\dist - \MSD^\cent | \le O(\nu^{1 + \gamma_o})
\ee
where both $\MSD^\dist$ and $\MSD^\cent$ are in the order of $\nu$.
\end{lemma}
\begin{IEEEproof}
From \eqref{eqn:Hcdef}, \eqref{eqn:Rcdef}, and \eqref{eqn:piandp}, it is easy to verify that $H_\c = H$ and $R_\c = R$, where $\{H, R\}$ are given by \eqref{II-eqn:Hdef} and \eqref{II-eqn:Rdef} of Part II \cite{Zhao13TSPasync2}. Using \eqref{eqn:MSDbatchapprox} and \eqref{eqn:MSDdiffusion} then completes the proof.
\end{IEEEproof}

\section{Comparison II: Asynchronous vs. Synchronous Networks}

Synchronous diffusion networks run \eqref{I-eqn:atcsync1}--\eqref{I-eqn:atcsync2} from Part I \cite{Zhao13TSPasync1}, namely,
\begin{subequations}
\begin{alignat}{2}
\label{eqn:atcsync1}
{\bm{\psi}}_{k,i} & = {\bm{w}}_{k,i-1} - \mu_k \wh{\nabla_{w^*}J_k}(\bm{w}_{k,i-1}) & \quad & \mbox{(adaptation)}\\
\label{eqn:atcsync2}
{\bm{w}}_{k,i} & = \sum_{\ell\in{\Ncal}_k}a_{\ell k}\,{\bm{\psi}}_{\ell,i} & \quad & \mbox{(combination)}
\end{alignat}
\end{subequations}
These networks can be viewed as a special case of asynchronous networks running \eqref{eqn:atcasync1}--\eqref{eqn:atcasync2} when the random step-sizes and combination coefficients assume constant values. If we set the covariances $\{ c_{\mu,k,k} \}$ and $\{ c_{a,\ell k, \ell k} \}$ to zero, then the asynchronous network \eqref{eqn:atcasync1}--\eqref{eqn:atcasync2} will reduce to the synchronous network \eqref{eqn:atcsync1}--\eqref{eqn:atcsync2} with the parameters $\{ \mu_k, a_{\ell k}\}$ replaced by $\{ \bar{\mu}_k, \bar{a}_{\ell k} \}$. We can therefore specialize the results obtained for asynchronous networks to the synchronous case by using $\{ \bar{\mu}_k \}$ and $\{ \bar{a}_{\ell k} \}$ and assuming $c_{\mu,k,k} = 0$ and $c_{a,\ell k,\ell k} = 0$ for all $k$ and $\ell$. For example, it is easy to verify that the mean error recursion of the long term model for the synchronous solution with $\{ \bar{\mu}_k \}$ and $\{ \bar{a}_{\ell k} \}$ is identical to \eqref{II-eqn:meanerrorrecursion} from Part II \cite{Zhao13TSPasync2} for the asynchronous solution.

Under Assumption \ref{asm:smallstepsizes}, the asynchronous network with the random parameters $\{ \bm{\mu}_k(i) \}$ and $\{ \bm{a}_{\ell k}(i) \}$ and the synchronous network with the constant parameters $\{ \bar{\mu}_k \}$ and $\{ \bar{a}_{\ell k} \}$ have similar mean-square convergence rates for large $i$, but the steady-state MSD performance of the former is larger than that of the latter by a small amount. This result is established as follows. From Theorem \ref{II-theorem:errorcovariancerecursion} in Part II \cite{Zhao13TSPasync2}, the mean-square convergence rate for the asynchronous network with large $i$ is determined by $\rho(\Fcal_{\async})$ where
\be
\label{eqn:bigFasyncdef}
\Fcal_{\async} = 
\E ( \bm{\Bcal}_i^\T \bkron \bm{\Bcal}_i^*)
\ee
and $\bm{\Bcal}_i$ is given by \eqref{II-eqn:bigBidef} of Part II \cite{Zhao13TSPasync2}. We are adding the subscript ``async'' to quantities that are related to asynchronous networks. Correspondingly, the mean-square convergence rate for the synchronous network with the constant parameters $\{ \bar{\mu}_k \}$ and $\{ \bar{a}_{\ell k} \}$ will be determined by $\rho(\Fcal_{\sync})$ where
\be
\label{eqn:bigFsyncdef}
\Fcal_{\sync} \defeq \bar{\Bcal}^\T \bkron \bar{\Bcal}^*
\ee
and $\bar{\Bcal}$ is given by \eqref{II-eqn:bigBmeandef} of Part II \cite{Zhao13TSPasync2}.

\begin{lemma}[{Matching mean-square convergence rates}]
\label{corollary:comparisonrates}
For large $i$, the mean-square convergence rate of the asynchronous diffusion strategy is close to that of the synchronous diffusion strategy:
\be
\label{eqn:rhobigFasyncapprox}
| \rho(\Fcal_{\async}) - \rho(\Fcal_{\sync}) | = O(\nu^{1+1/N^2})
\ee
where $\rho(\Fcal_{\async})$ and $\rho(\Fcal_{\sync})$ are both dominated by $[1 - \lambda_{\min}(H)]^2 = 1 - O(\nu)$ for small $\nu$ by Assumption \ref{asm:smallstepsizes}.
\end{lemma}

\begin{IEEEproof}
By Lemma \ref{II-lemma:lowrank} of Part II \cite{Zhao13TSPasync2}, we have
\be
\label{eqn:rhoFasyncorder}
\rho(\Fcal_{\async}) = \rho(F_{\async}) + O(\nu^{1+1/N^2})
\ee
where $F_{\async}$ is given by \eqref{II-eqn:Fdef} of Part II \cite{Zhao13TSPasync2}. Correspondingly, we will also have
\be
\label{eqn:rhoFsyncorder}
\rho(\Fcal_{\sync}) = \rho(F_{\sync}) + O(\nu^{1+1/N^2})
\ee
where $F_{\sync}$ is given by
\be
\label{eqn:Fsyncdef}
F_{\sync} = \sum_{k = 1}^{N}\sum_{\ell = 1}^{N} \bar{p}_{\ell} \bar{p}_{k} ( \bar{D}_\ell^\T\kron \bar{D}_k )
\ee
Noting that $F_{\sync}$ is identical to $F'$ in \eqref{II-eqn:Fpdef} of Part II \cite{Zhao13TSPasync2}, then from Lemma \ref{II-lemma:spectralF} of Part II \cite{Zhao13TSPasync2} we obtain
\be
\label{eqn:rhoFasyncclosetorhoFsync}
\rho(F_{\async}) = \rho(F_{\sync}) + O(\nu^2)
\ee
Using \eqref{eqn:rhoFasyncorder}, \eqref{eqn:rhoFsyncorder}, and \eqref{eqn:rhoFasyncclosetorhoFsync}, we get
\begin{align}
\label{eqn:approxrhos}
| \rho(\Fcal_{\async}) - \rho(\Fcal_{\sync}) | & = | \rho(F_{\async}) - \rho(F_{\sync}) + O(\nu^{1+1/N^2}) | \nn \\
& = | O(\nu^2) + O(\nu^{1+1/N^2}) | \nn \\
& = O(\nu^{1+1/N^2})
\end{align}
Using \eqref{II-eqn:rhoFapprox} from Part II \cite{Zhao13TSPasync2} and \eqref{eqn:approxrhos} completes the proof.
\end{IEEEproof}

Likewise, assuming $c_{\mu,k,k}  =  0$ and $c_{a,\ell k,\ell k}  =  0$ for all $k$ and $\ell$ for the synchronous strategy, it is easy to verify from \eqref{II-eqn:pdef}--\eqref{II-eqn:Ppdef} of Part II \cite{Zhao13TSPasync2} that $p  =  \bar{p} \kron \bar{p}$. Then, we obtain the following expression for the steady-state MSD of the synchronous network with the constant parameters $\{ \bar{\mu}_k \}$ and $\{ \bar{a}_{\ell k} \}$:
\be
\label{eqn:steadyMSDapprox_const}
\MSD_{\sync}^{\dist} = \frac{1}{4} \Tr( H^{-1} R_{\sync}) + O(\nu^{1 + \gamma_o})
\ee
where $H = O(\nu)$ and $0 < \gamma_o \le 1/2$ are given by \eqref{II-eqn:Hdef} and \eqref{II-eqn:gammaodef} from Part II \cite{Zhao13TSPasync2}, respectively, and
\be
\label{eqn:Rsyncdef}
R_{\sync} \defeq \sum_{k=1}^{N} \bar{p}_k^2 \, \bar{\mu}_k^2 R_k = O(\nu^2)
\ee
Since $\Tr( H^{-1} R_{\sync}) = O(\nu)$, the first term on the RHS of \eqref{eqn:steadyMSDapprox_const} dominates the other term, $O(\nu^{1 + \gamma_o})$. From \eqref{eqn:MSDdiffusion} and \eqref{eqn:steadyMSDapprox_const}, we observe that the network MSDs of asynchronous and synchronous networks are both in the order of $\nu$.

\begin{lemma}[{Degradation in MSD is $O(\nu)$}]
\label{corollary:comparison}
The network MSD \eqref{eqn:MSDdiffusion} for the asynchronous  diffusion strategy is greater than the network MSD \eqref{eqn:steadyMSDapprox_const} for the synchronous diffusion strategy by a difference in the order of $\nu$.
\end{lemma}
\begin{IEEEproof}
The difference between $R_{\async}$ and $R_{\sync}$ is
\begin{align}
\label{eqn:compareMSDpart2}
R_{\async} - R_{\sync} = \sum_{k=1}^{N} [ (p_{k,k} - \bar{p}_k^2) \bar{\mu}_k^2 + p_{k,k} c_{\mu,k,k}] R_k
\end{align}
where $R_{\async}$ is given by \eqref{II-eqn:Rdef} of Part II \cite{Zhao13TSPasync2}. Since $p_{k,k} - \bar{p}_k^2$ is the $k$-th entry on the diagonal of $P_p - \bar{p}\bar{p}^\T$, from Lemma \ref{lemma:nonnegativity}, we know that all entries on the diagonal of $P_p - \bar{p}\bar{p}^\T$ are nonnegative, which implies that $p_{k,k} - \bar{p}_k^2 \ge 0$. Moreover, by Perron-Frobenius Theorem \cite{BermanPF}, all entries of the Perron eigenvector $p$ must be positive, which implies that $p_{k,k} > 0$. We also know that $c_{\mu,k,k}$ must be positive in the asynchronous model. Therefore, we get
\be
(p_{k,k} - \bar{p}_k^2) \bar{\mu}_k^2 + p_{k,k} c_{\mu,k,k} > 0
\ee
Moreover, by using \eqref{II-eqn:bigMbound}--\eqref{II-eqn:bigCMbound2} from Part II \cite{Zhao13TSPasync2}, we have
\be
\label{eqn:pkkminuspbark2Onu2}
(p_{k,k} - \bar{p}_k^2) \bar{\mu}_k^2 + p_{k,k} c_{\mu,k,k} = O(\nu^2)
\ee
Then, using the fact that the $\{R_k\}$ are positive semi-definite, we conclude from \eqref{eqn:compareMSDpart2}--\eqref{eqn:pkkminuspbark2Onu2} that
\be
\| R_{\async} - R_{\sync} \| = O(\nu^2) > 0
\ee
From \eqref{II-eqn:Hdef} of Part II \cite{Zhao13TSPasync2}, we know that $H^{-1} = O(\nu^{-1})$. Therefore, we get
\begin{align}
\MSD_{\async}^{\dist} - \MSD_{\sync}^{\dist} & = \frac{1}{4}\Tr[H^{-1}(R_{\async}  -   R_{\sync})]  +  O(\nu^{1 + \gamma_o}) \nn \\
& = O(\nu) + O(\nu^{1 + \gamma_o}) = O(\nu)
\end{align}
and
\be
\MSD_{\async}^{\dist} - \MSD_{\sync}^{\dist} \ge 0
\ee
which complete the proof.
\end{IEEEproof}

We observe from the above results that when the step-sizes are sufficiently small, the mean-square convergence rate of the asynchronous network tends to be immune from the uncertainties caused by random topologies, links, agents, and data arrival time. However, there is an $O(\nu)$ degradation in the steady-state MSD level for the asynchronous network -- refer to Table \ref{table:summary} for a summary of the main conclusions.

\section{A Case Study: MSE Estimation}
\label{sec:LMS}

The previous results apply to arbitrary strongly-convex costs $\{J_k(w)\}$ whose Hessian functions are locally Lipschitz continuous at $w^o$. In this section we specialize the results to the case of MSE estimation over networks, where the costs $\{J_k(w)\}$ become quadratic in $w \in \mbbC^{M \times 1}$.

\subsection{Problem Formulation and Modeling}
We now assume that each agent $k$ has access to streaming data $\{\bm{d}_k(i),\bm{u}_{k,i}\}$ related via the linear regression model:
\be
\label{eqn:linearmodel}
\bm{d}_k(i) = \bm{u}_{k,i} w^o + \bm{\xi}_k(i)
\ee
where $\bm{d}_k(i)\in\mbbC$ is the observation, $\bm{u}_{k,i}\in\mbbC^{1\times M}$ is the regressor, $w^o\in\mbbC^{M\times 1}$ is the desired parameter vector, and $\bm{\xi}_k(i)$ is additive noise. 

\begin{assumption}[Data model]
\label{asm:datamodel}
\hfill
\begin{enumerate}
\item The regressors $\{\bm{u}_{k,i}\}$ are temporally white and spatially independent circular symmetric complex random variables with zero mean and covariance matrix $R_{u,k} > 0$.
\item The noise signals $\{\bm{\xi}_k(i)\}$ are temporally white and spatially independent circular symmetric complex random variables with zero mean and variance $\sigma_{\xi,k}^2 > 0$. 
\item The random variables $\{\bm{u}_{k,i}, \bm{\xi}_\ell(j)\}$ are mutually independent for any $k$ and $\ell$, $i$ and $j$, and they are independent of any other random variable. 
\end{enumerate}
\end{assumption}

\noindent The objective for the network is to estimate $w^o$ by minimizing the aggregate mean-square-error cost defined by
\be
\minimize_{w} \;\; \sum_{k=1}^{N} J_k(w) \defeq \sum_{k=1}^{N}\E\,|\bm{d}_k(i) - \bm{u}_{k,i} w|^2
\ee
It can be verified that this problem satisfies Assumptions \ref{I-asm:costfunctions} and \ref{I-asm:boundedHessian} introduced in Part I \cite{Zhao13TSPasync1}.

\subsection{Distributed Diffusion Solutions}
The asynchronous diffusion solution \eqref{eqn:atcasync1}--\eqref{eqn:atcasync2} will then reduce to the following form:
\begin{subequations}
\begin{align}
\label{eqn:atcasync1lms}
\bm{\psi}_{k,i} & = \bm{w}_{k,i-1} + \bm{\mu}_k(i) \bm{u}_{k,i}^* [\bm{d}_k(i) - \bm{u}_{k,i} \bm{w}_{k,i-1}] \\
\label{eqn:atcasync2lms}
\bm{w}_{k,i} & = \sum_{\ell \in \bm{\Ncal}_{k,i}} \bm{a}_{\ell k}(i) \,\bm{\psi}_{\ell,i}
\end{align}
\end{subequations}
and the synchronous network \eqref{eqn:atcsync1}--\eqref{eqn:atcsync2} will become
\begin{subequations}
\begin{align}
\label{eqn:atcsync1lms}
{\bm{\psi}}_{k,i} & = {\bm{w}}_{k,i-1} + \bar{\mu}_k\bm{u}_{k,i}^*[\bm{d}_k(i) - \bm{u}_{k,i}{\bm{w}}_{k,i-1}] \\
\label{eqn:atcsync2lms}
{\bm{w}}_{k,i} & = \sum_{\ell\in{\Ncal}_{k,i}} \bar{a}_{\ell k}\,{\bm{\psi}}_{\ell,i}
\end{align}
\end{subequations}
We assume that the network is under the Bernoulli model described in Part I \cite{Zhao13TSPasync1}. For illustration purposes only, we assume that the parameters $\{\mu_k\}$ in \eqref{I-eqn:binary_randomstepsize} of Part I \cite{Zhao13TSPasync1} are uniform, $\mu_k \equiv \mu$, and that the parameters $\{a_{\ell k}; \ell \in \Nknk\}$ in \eqref{I-eqn:binary_randomcombine} of Part I \cite{Zhao13TSPasync1} are given by $a_{\ell k} = |\Ncal_k|^{-1}$.

Substituting \eqref{eqn:linearmodel} into \eqref{eqn:atcasync1lms} and comparing with \eqref{I-eqn:linearperturbationmodel} of Part I \cite{Zhao13TSPasync1}, we find that the approximate gradient, $\wh{\nabla_{w^*}J_k}(\bm{w}_{k,i-1})$, and the corresponding gradient noise, $\bm{v}_{k,i}(\bm{w}_{k,i-1})$, in this case are given by
\begin{align}
\label{eqn:gradient_lms}
\wh{\nabla_{w^*}J_k}(\bm{w}_{k,i-1}) & = - \bm{u}_{k,i}^*[\bm{d}_k(i) - \bm{u}_{k,i}{\bm{w}}_{k,i-1}] \nn \\
& = - \bm{u}_{k,i}^*\bm{u}_{k,i}\wt{\bm{w}}_{k,i-1} - \bm{u}_{k,i}^*\bm{\xi}_k(i) \nn \\
& = - R_{u,k}\wt{\bm{w}}_{k,i-1} - \bm{v}_{k,i}(\bm{w}_{k,i-1})
\end{align}
where
\be
\label{eqn:gradientnoise_lms}
\bm{v}_{k,i}(\bm{w}_{k,i-1}) = (\bm{u}_{k,i}^*\bm{u}_{k,i}  -  R_{u,k}) \wt{\bm{w}}_{k,i-1} + \bm{u}_{k,i}^*\bm{\xi}_k(i)
\ee
It can be verified that the gradient noise $\bm{v}_{k,i}(\bm{w}_{k,i-1})$ in \eqref{eqn:gradientnoise_lms} satisfies Assumption \ref{II-asm:gradientnoisestronger} of Part II \cite{Zhao13TSPasync2} and that the covariance matrix of $\ubar{\bm{v}}_{k,i}(w^o) = \ubar{\mbbT}( \bm{u}_{k,i}^*\bm{\xi}_k(i) )$, where $\ubar{\mbbT}(\cdot)$ is from \eqref{I-eqn:ubardef} of Part I \cite{Zhao13TSPasync1}, is given by
\be
\label{eqn:noisecov_lms}
R_k = \diag\{\sigma_{\xi,k}^2 R_{u,k}, \sigma_{\xi,k}^2 R_{u,k}^\T \} \defeq \sigma_{\xi,k}^2 H_k
\ee
Moreover, the complex Hessian of the cost $J_k(w)$ is given by
\be
\label{eqn:hessian_lms}
\nabla_{\ubar{w}\ubar{w}^*}^2 J_k(\ubar{w}) \defeq H_k =
\diag\{R_{u,k}, R_{u,k}^\T \}
\ee
We further note that for the Bernoulli network under study,
\begin{align}
\bar{\mu}_k^{(1)} = q_k \mu, \quad \bar{\mu}_k^{(2)} = q_k \mu^2
\end{align}
Therefore, the parameter $\nu = \mu$ in this case. If $\mu$ is small enough and satisfies Assumption \ref{asm:smallstepsizes}, then from \eqref{eqn:MSDdiffusion}, the network MSD of the asynchronous network is given by
\begin{align}
\label{eqn:MSDLMS}
\MSD_{\async}^{\diff} =  \frac{\mu}{2} \Tr    \left[  \left( \sum_{k = 1}^{N} \bar{p}_{k} q_k R_{u,k} \right)^{-1}    \left( \sum_{k=1}^{N} p_{k,k} q_k \sigma_{\xi,k}^2 R_{u,k} \right)   \right] + O(\mu^{1 + \gamma_o })
\end{align}
Likewise, the network MSD of the synchronous network from \eqref{eqn:steadyMSDapprox_const} is given by
\begin{align}
\label{eqn:MSDLMS_const}
\MSD_{\sync}^{\diff} &  =  \frac{\mu}{2} \Tr    \left[   \left( \sum_{k = 1}^{N} \bar{p}_k q_k R_{u,k} \right)^{-1}    \left( \sum_{k=1}^{N} \bar{p}_k^2 q_k^2 \sigma_{\xi,k}^2 R_{u,k} \right) \right] + O(\mu^{1 + \gamma_o})
\end{align}
Clearly, since $q_k \le 1$ and $p_{k,k}\ge \bar{p}_k^2$ for all $k$, the MSD in \eqref{eqn:MSDLMS} is always greater than the MSD in \eqref{eqn:MSDLMS_const} and the difference is in the order of $\mu$.

\subsection{Centralized Solution}
The asynchronous batch solution \eqref{eqn:batchasyn} will now reduce to
\be
\label{eqn:batchasynclms}
{\bm{w}}_{\c,i}  =  {\bm{w}}_{\c,i-1}  +  \sum_{k=1}^{N} \bm{\pi}_{k}(i) \bm{\mu}_k(i) \bm{u}_{k,i}^*[\bm{d}_k(i)  -  \bm{u}_{k,i}{\bm{w}}_{\c,i-1}]  
\ee
and the synchronous batch solution \eqref{eqn:batchsyn} will become
\be
\label{eqn:batchsynclms}
{\bm{w}}_{\c,i} = {\bm{w}}_{\c,i-1} + \sum_{k=1}^{N} \bar{\pi}_{k} \bar{\mu}_k \bm{u}_{k,i}^*[\bm{d}_k(i) - \bm{u}_{k,i}{\bm{w}}_{\c,i-1}]
\ee
We continue to assume that the random step-size parameters $\{\bm{\mu}_k(i)\}$ satisfy the same Bernoulli model described in Part I \cite{Zhao13TSPasync1} with a uniform profile $\mu_k \equiv \mu$. We use the procedure described in Section \ref{subsec:construct} to generate the random fusion coefficients $\{ \bm{\pi}_k(i) \}$. Specifically, we have $\bm{\pi}_k(i) = \bm{\phi}_k(i)$, where $\bm{\phi}_k(i)$ denotes the $k$-th entry of $\bm{\phi}_i$ from \eqref{eqn:phidef}.

\subsection{Simulation Results}

\begin{figure}
\centering
\includegraphics[scale=1]{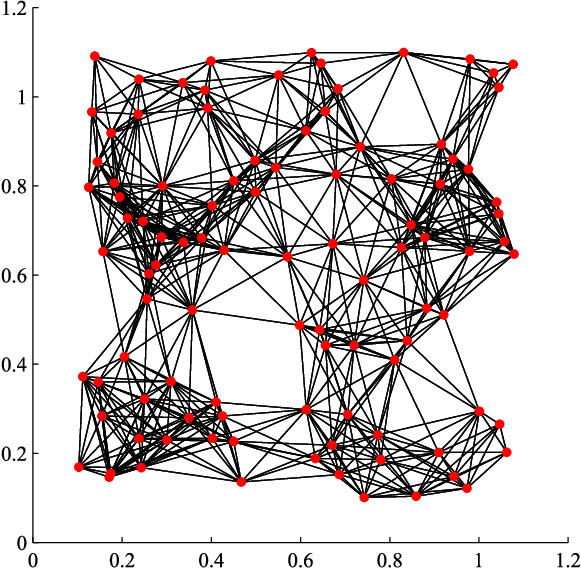}
\caption{A topology with 100 nodes.}
\label{fig:topology}
\vspace{-1\baselineskip}
\end{figure}

\begin{figure}
\centering
\includegraphics[scale=0.45]{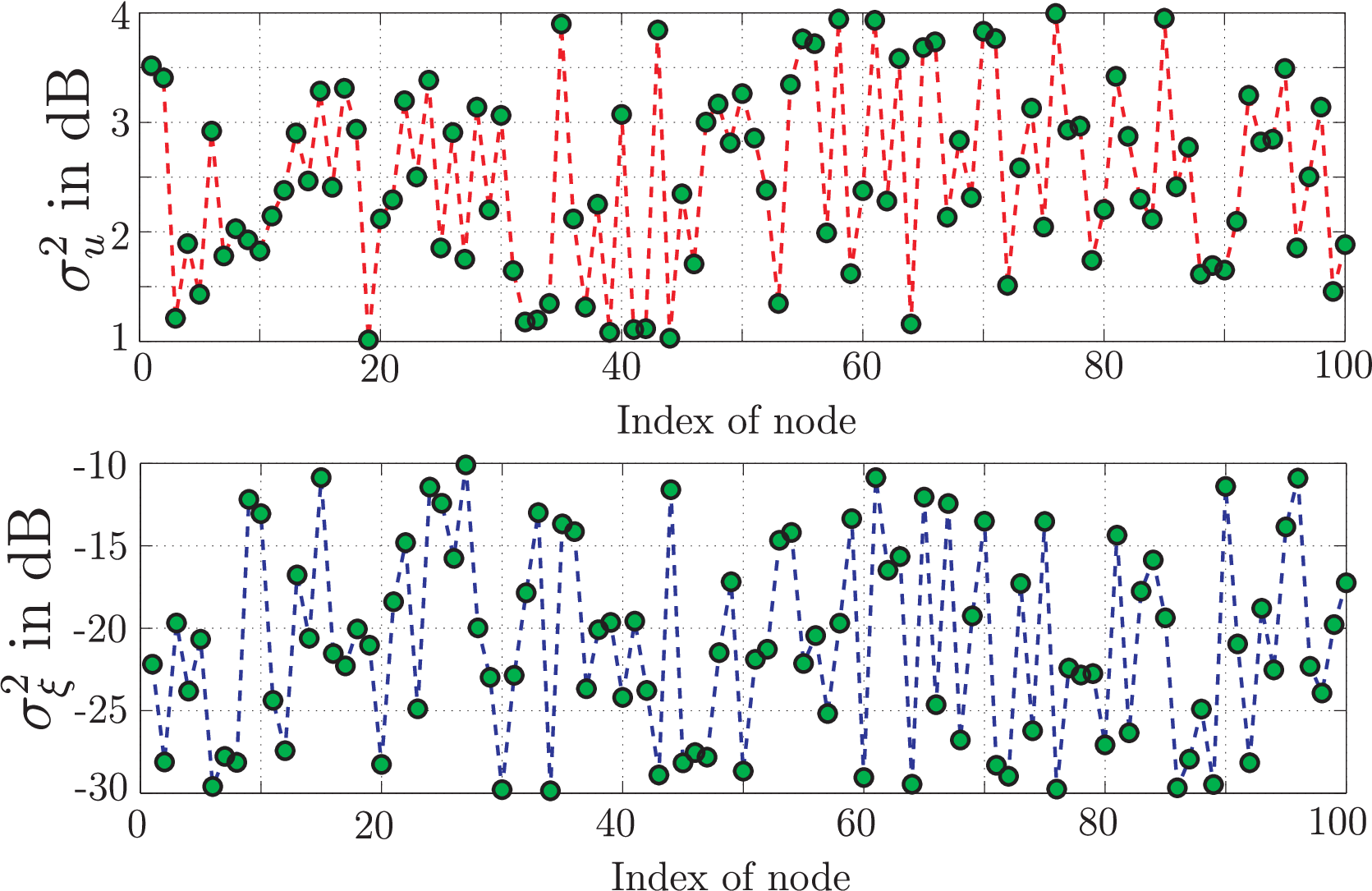}
\caption{Values of $\{\sigma_{u,k}^2\}$ and $\{\sigma_{\xi,k}^2\}$.}
\label{fig:profile}
\vspace{-1\baselineskip}
\end{figure}

\begin{figure}
\centering
\includegraphics[scale=0.8]{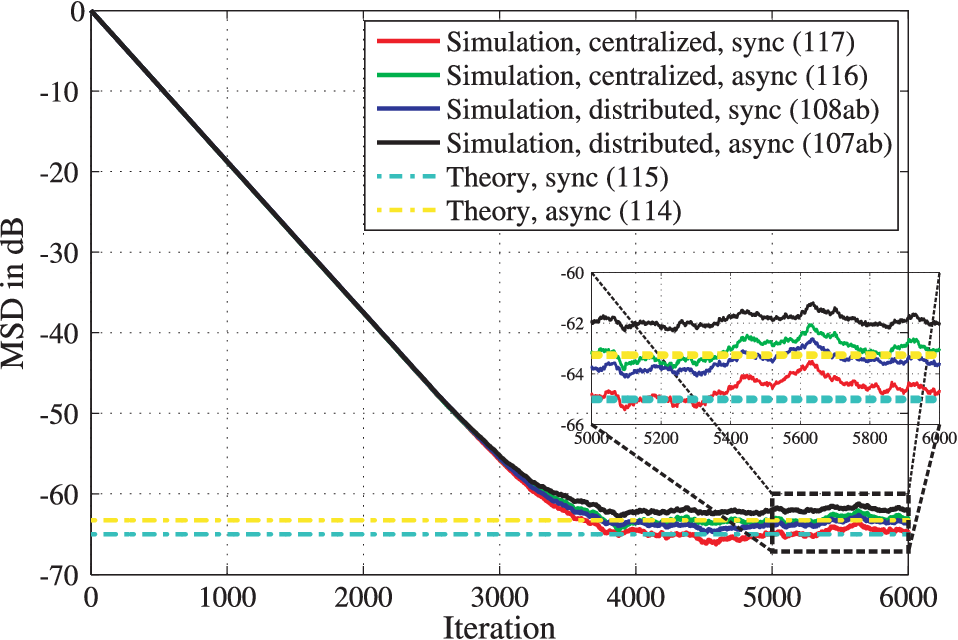}
\caption{MSD learning curves for the asynchronous and synchronous modes of operation.}
\label{fig:MSD}
\vspace{-1\baselineskip}
\end{figure}

We consider a network consisting of $N = 100$ agents with the connected topology shown in Fig.~\ref{fig:topology} where each link is assumed to be bidirectional. The length of the unknown parameter $w^o$ is set to $M = 2$. The regressors are assumed to be white, i.e., $R_{u,k} = \sigma_{u,k}^2 I_M$.  The values of $\{\sigma_{u,k}^2, \sigma_{v,k}^2\}$ are randomly generated and shown in Fig.~\ref{fig:profile}. The step-size parameter is set to $\mu = 0.002$. We randomly select the values for the probabilities $\{\eta_{\ell k}\}$ in \eqref{I-eqn:binary_randomcombine} of Part I \cite{Zhao13TSPasync1} within the range $(0.4,0.8)$, and randomly select the values for the probabilities $\{q_k\}$ in \eqref{I-eqn:binary_randomstepsize} of Part I \cite{Zhao13TSPasync1} within the set $\{0.3, 0.5, 0.7, 0.9\}$. The asynchronous distributed strategy \eqref{eqn:atcasync1lms}--\eqref{eqn:atcasync2lms}, the synchronous distributed strategy \eqref{eqn:atcsync1lms}--\eqref{eqn:atcsync2lms}, the asynchronous centralized solution \eqref{eqn:batchasynclms}, and the synchronous centralized solution \eqref{eqn:batchsynclms} are all simulated over 100 trials and 6000 iterations for each trial. The random fusion coefficients $\{\bm{\pi}_k(i)\}$ are obtained by sampling $\bm{\phi}_i$ from \eqref{eqn:phidef}. The $\bm{\phi}_i$ is constructed by consecutively multiplying 100 independent realizations of $\bm{A}_i$. The averaged learning curves (MSD) as well as the theoretical MSD results \eqref{eqn:MSDLMS} for asynchronous solutions and \eqref{eqn:MSDLMS_const} for synchronous solutions are plotted in Fig.~\ref{fig:MSD}. We observe a good match between theory and simulation. We also observe that both synchronous and asynchronous solutions converge at a similar rate but that the former attains a lower MSD level at steady-state as predicted by \eqref{eqn:MSDLMS} and \eqref{eqn:MSDLMS_const}.

\section{Conclusion}
In this part, we compared the performance of distributed and centralized solutions under two modes of operation: synchronous and asynchronous implementations. We derived explicit comparisons for the mean and mean-square rates of convergence, as well as for the steady-state mean-square error performance. The main results are captured by Table 1. It is seen that diffusion networks are remarkably resilient to asynchronous or random failures: the convergence continues to occur at the same rate as synchronous or centralized solutions while the MSD level suffers a degradation in the order of $O(\nu)$ relative to synchronous diffusion networks. The results in the article highlight yet another benefit of cooperation: remarkable resilience to random failures and asynchronous events.

\appendices

\section{Proof of Lemma \ref{lemma:propertiesFc}}
\label{app:propertiesFc}
From \eqref{eqn:Bidef}, we get
\begin{align}
\label{eqn:Fcexpressionproof}
F_\c & = \E\left[ \left( \sum_{\ell=1}^{N} \bm{\pi}_{\ell}(i) \bm{D}_{\ell,i} \right)^\T \kron \left( \sum_{k=1}^{N} \bm{\pi}_{k}(i) \bm{D}_{k,i} \right) \right] \nn \\
& \stackrel{(a)}{=} \sum_{k=1}^{N} \sum_{\ell=1}^{N} \E[\bm{\pi}_{\ell}(i)\bm{\pi}_{k}(i) ] \cdot \E (  \bm{D}_{\ell,i}^\T \kron   \bm{D}_{k,i} ) \nn \\
& \stackrel{(b)}{=} \sum_{k=1}^{N} \sum_{\ell=1}^{N} (\bar{\pi}_\ell \bar{\pi}_k  +  c_{\pi,\ell,k})(\bar{D}_\ell^\T \kron \bar{D}_k  +  c_{\mu,\ell,k} H_\ell^\T \kron H_k)   
\end{align}
where step (a) is by using the independence condition from the asynchronous model; and step (b) is by using \eqref{eqn:randstepsizemeanentry}--\eqref{eqn:randcombinecoventry}. Since $\{\bar{D}_k, H_k\}$ are all Hermitian, it is straightforward to verify that $F_\c$ is also Hermitian.

Using Jensen's inequality and the convexity of the $2$-induced norm, $\|\cdot\|$, we obtain from \eqref{eqn:Fcdef} that
\be
\label{eqn:rhoFcbound1}
\rho(F_\c) \le \E\,\| \bm{B}_i^\T \kron \bm{B}_i \| = \E\,\| \bm{B}_i \|^2
\ee
where we used the identities $\| A \kron B\| = \|A\| \cdot \|B\|$ \cite[p.~245]{Horn91} and $\| A^\T \| = \| A \|$. 
Using Jensen's inequality again with respect to the convex coefficients $\{\bm{\pi}_k(i)\}$ and the fact that $\| \cdot \|^2$ is also a convex function, we get from \eqref{eqn:Bidef} that
\be
\label{eqn:boundnorm2Bi}
\| \bm{B}_i \|^2 = \left\| \sum_{k=1}^{N} \bm{\pi}_{k}(i) \bm{D}_{k,i} \right\|^2 \le \sum_{k=1}^{N} \bm{\pi}_{k}(i) \| \bm{D}_{k,i} \|^2
\ee
Substituting \eqref{eqn:boundnorm2Bi} into \eqref{eqn:rhoFcbound1}, we obtain
\begin{align}
\label{eqn:rhoFcbound2}
\rho(F_\c) \le \sum_{k=1}^{N} \bar{\pi}_k \E\| \bm{D}_{k,i} \|^2 \le \max_k \E\| \bm{D}_{k,i} \|^2
\end{align}
From \eqref{eqn:Bidef} and from condition \eqref{I-eqn:boundseigHessian} in Part I \cite{Zhao13TSPasync1}, we have
\be
\label{eqn:boundDkieig}
1 - \bm{\mu}_k(i) \lambda_{k,\max} \le \lambda(\bm{D}_{k,i}) \le 1 - \bm{\mu}_k(i) \lambda_{k,\min}
\ee
for every eigenvalue of $\bm{D}_{k,i}$ and for every $k$ and $i\ge0$. Since $\bm{D}_{k,i}$ is Hermitian, we conclude from \eqref{eqn:boundDkieig} that for every $k$ and $i\ge0$, 
\begin{align}
\label{eqn:boundDkinorm}
\| \bm{D}_{k,i} \|^2 & \le \max\{ [1 - \bm{\mu}_k(i) \lambda_{k,\min}]^2, [1 - \bm{\mu}_k(i) \lambda_{k,\max}]^2 \} \nn \\
& \le 1 - 2 \bm{\mu}_k(i) \lambda_{k,\min} + \bm{\mu}_k^2(i) \lambda_{k,\max}^2
\end{align}
Substituting \eqref{eqn:boundDkinorm} into \eqref{eqn:rhoFcbound2} yields
\begin{align}
\label{eqn:rhoFcbound3}
\rho(F_\c) & \le \max_k \E\,[ 1 - 2 \bm{\mu}_k(i) \lambda_{k,\min} + \bm{\mu}_k^2(i) \lambda_{k,\max}^2 | \ubar{\wt{\bm{w}}}_{\c,i-1} ] \nn \\
& \le \max_k \{ 1 - 2 \bar{\mu}_k \lambda_{k,\min} + (\bar{\mu}_k^2 + c_{\mu,k,k}) \lambda_{k,\max}^2 \} \nn \\
& < \max_k\{ \gamma_k^2 + \alpha (\bar{\mu}_k^2 + c_{\mu,k,k}) \} \nn \\
& = \beta
\end{align}
where $\alpha > 0$ and $\{\gamma_k^2, \beta\}$ are from \eqref{I-eqn:gammak2def} and \eqref{I-eqn:betadef} of Part I \cite{Zhao13TSPasync1}, respectively. In \eqref{I-eqn:meansquarestabilitycond2} from Part I \cite{Zhao13TSPasync1}, we established that $|\beta| < 1$ if condition \eqref{eqn:meansquarestablecondition} holds. Therefore, by \eqref{eqn:rhoFcbound3} we conclude that $\rho(F_\c) < 1$ when condition \eqref{eqn:meansquarestablecondition} holds.

Since $c_{\mu,\ell,k} = O(\nu^2)$ by using \eqref{II-eqn:bigCMbound1} and \eqref{II-eqn:bigCMbound2} from Part II \cite{Zhao13TSPasync2}, we get from \eqref{eqn:Fcexpressionproof} and \eqref{eqn:Bmeandef} that
\be
\label{eqn:Fcorderpart1}
F_\c = \sum_{k=1}^{N} \sum_{\ell=1}^{N} (\bar{\pi}_\ell \bar{\pi}_k + c_{\pi,\ell,k}) (\bar{D}_\ell^\T \kron \bar{D}_k) + O(\nu^2)
\ee
Furthermore, we have
\begin{align}
\label{eqn:Fcorderpart2}
\sum_{k=1}^{N} \sum_{\ell=1}^{N} c_{\pi,\ell,k} (\bar{D}_\ell^\T \kron \bar{D}_k) 
& \stackrel{(a)}{=} \sum_{k=1}^{N} \sum_{\ell=1}^{N} c_{\pi,\ell,k} (I_{2M} - \bar{\mu}_\ell H_\ell)^\T \kron (I_{2M} - \bar{\mu}_k H_k) \nn \\
& = \sum_{k=1}^{N} \sum_{\ell=1}^{N} c_{\pi,\ell,k} (I_{4M^2} - \bar{\mu}_\ell H_\ell^\T \kron I_{2M} - I_{2M} \kron \bar{\mu}_k H_k + \bar{\mu}_\ell \bar{\mu}_k H_\ell^\T \kron H_k) \nn \\
& \stackrel{(b)}{=} \sum_{k=1}^{N} \sum_{\ell=1}^{N} c_{\pi,\ell,k} \bar{\mu}_\ell \bar{\mu}_k (H_\ell^\T \kron H_k) \nn \\
& \stackrel{(c)}{=} O(\nu^2)
\end{align}
where step (a) is by using \eqref{eqn:Dkmeandef}; step (b) is by using \eqref{eqn:momentspikicondtions}; and step (c) is by using \eqref{II-eqn:bigCMbound1} and \eqref{II-eqn:bigCMbound2} from Part II \cite{Zhao13TSPasync2}. From \eqref{eqn:Fcorderpart1} and \eqref{eqn:Fcorderpart2}, we have
\be
\label{eqn:Fcapprox}
F_\c = \sum_{\ell = 1}^{N} \sum_{k = 1}^{N} \bar{\pi}_\ell \bar{\pi}_k (\bar{D}_\ell^\T \kron \bar{D}_k) + O(\nu^2)
\ee
Now, consider the matrix $F_\c'$ defined in \eqref{eqn:Fcprimedef}; it is easy to verify by using \eqref{eqn:Bmeandef} that
\be 
\label{eqn:Fcprimeexpress}
F_\c' = \sum_{\ell = 1}^{N} \sum_{k = 1}^{N} \bar{\pi}_\ell \bar{\pi}_k (\bar{D}_\ell^\T \kron \bar{D}_k) = \bar{B}^\T \kron \bar{B}
\ee
Since $\bar{B}$ is Hermitian, so is $F_\c'$. From \eqref{eqn:Fcapprox} and \eqref{eqn:Fcprimeexpress}, we get $\| F_\c - F_\c' \| = O(\nu^2)$. Since both $F_\c$ and $F_\c'$ are Hermitian, their difference $F_\c - F_\c'$ is also Hermitian. Then, using a corollary of the Wielandt-Hoffman Theorem \cite{Golub96}, we conclude that
\be
\label{eqn:lambdaFcandlambdaFcprime}
| \lambda_m(F_\c) - \lambda_m(F_\c') | \le  \| F_\c - F_\c' \| = O(\nu^2)
\ee
where $\lambda_m(\cdot)$ denotes the $m$-th eigenvalue of its Hermitian matrix argument; the eigenvalues are assumed to be ordered from largest to smallest in each case. From \eqref{eqn:lambdaFcandlambdaFcprime}, we immediately deduce that
\be
\label{eqn:rhoFcandrhoFcprime}
|\rho(F_\c) - \rho(F_\c')| \le O(\nu^2)
\ee
From \eqref{eqn:Bmeandef}--\eqref{eqn:Dkmeandef} and \eqref{eqn:Hcdef}, we have
\be
\label{eqn:Bmeannew}
\bar{B} = I_{2M} - H_\c, \qquad \lambda(\bar{B}) = 1 - \lambda(H_\c)
\ee
Since $H_\c$ is symmetric positive definite, and since the $\{\bar{\pi}_k\}$ are convex coefficients by \eqref{eqn:momentspikicondtions}, we get from Jensen's inequality that
\be
\label{eqn:lambdaHcexpress}
0  <  \lambda(H_\c)  \le  \| H_\c \|  \le  \sum_{k=1}^{N} \bar{\pi}_k \|\bar{\mu}_k H_k \|  \le  \max_k \{ \bar{\mu}_k \lambda_{k,\max} \}  
\ee
for all eigenvalues of $H_\c$. When condition \eqref{eqn:meansquarestablecondition} holds, we have
\be
\bar{\mu}_k \le \bar{\mu}_k (1+\rho_k^2) < \frac{\lambda_{k,\min}}{\alpha + \lambda_{k,\max}^2} < \frac{1}{\lambda_{k,\max}}
\ee
for any $k$. This implies that $\max_k \{ \bar{\pi}_k \lambda_{k,\max} \} < 1$ and therefore, $0 < \lambda(H_\c) < 1$ for all eigenvalues of $H_\c$. From \eqref{II-eqn:bigMbound} of Part II \cite{Zhao13TSPasync2} and \eqref{eqn:lambdaHcexpress}, we get
\be
\label{eqn:eigenvalueHc}
0 < \lambda(H_\c) = O(\nu) < 1
\ee
for any eigenvalue of $H_\c$. Therefore, we get from \eqref{eqn:Bmeannew} that
\be
\label{eqn:rhobarBexpression}
\lambda(\bar{B}) = 1 - O(\nu), \quad \rho(\bar{B}) = 1 - \lambda_{\min}(H_\c)
\ee
Then, from \eqref{eqn:Fcprimeexpress} and \eqref{eqn:rhobarBexpression}, we have
\be
\label{eqn:rhoFcprimeexpression}
\rho(F_\c') = [1 - \lambda_{\min}(H_\c)]^2
\ee
It then follows from \eqref{eqn:rhoFcandrhoFcprime} and \eqref{eqn:rhoFcprimeexpression} that
\be
\label{eqn:rhoFcexpress}
\rho(F_\c) = [ 1- \lambda_{\min}(H_\c) ]^2 + O(\nu^2)
\ee
where $\lambda_{\min}(H_\c) = O(\nu)$. Under Assumption \ref{asm:smallstepsizes}, we have
\be
[ 1- \lambda_{\min}(H_\c) ]^2 = 1 - 2\lambda_{\min}(H_\c) + O(\nu^2) = 1 - O(\nu)
\ee
which therefore dominates the $O(\nu^2)$ in \eqref{eqn:rhoFcexpress}.

From \eqref{eqn:Bmeannew} and \eqref{eqn:Fcprimeexpress}, we get
\be
\label{eqn:Fcprimeexpand}
F_\c' = I_{4M^2} - H_\c^\T \kron I_{2M} - I_{2M} \kron H_\c + H_\c^\T \kron H_\c
\ee
Then, using \eqref{eqn:Fcapprox}, \eqref{eqn:Fcprimeexpress}, and \eqref{eqn:Fcprimeexpand}, we have
\be
\label{eqn:IminusFcexpression}
I_{4M^2} - F_\c = \underbrace{H_\c^\T \kron I_{2M} + I_{2M} \kron H_\c}_{ = \; O(\nu)} + O(\nu^2)
\ee
where we used the fact that $H_\c^\T \kron H_\c = O(\nu^2)$ since $H_\c = O(\nu)$ by \eqref{eqn:Hcdef}. Using the fact that $H_\c$ is positive definite and is of the order of $\nu$, we eventually get
\be
\| (I_{4M^2} - F_\c)^{-1} \| = O(\nu^{-1})
\ee

\section{Proof of Theorem \ref{theorem:MSD}}
\label{app:MSD}
We start with the $\lim_{i \rightarrow \infty} y_{c,i}$ in \eqref{eqn:zcinfdef}. From \eqref{eqn:ycidef}, we have
\be
\label{eqn:limiyci}
\lim_{i \rightarrow \infty} y_{c,i} = \lim_{i \rightarrow \infty} \vecm\left( \E \ubar{\bm{s}}_i \ubar{\bm{s}}_i^* \right)
\ee
Using the gradient noise model from Section \ref{II-asm:gradientnoisestronger} of Part II \cite{Zhao13TSPasync2}, it can be verified that $\ubar{\bm{s}}_i$ is zero mean and that its conditional covariance matrix is given by
\begin{align}
\label{eqn:sicovariance}
\E\,[ \ubar{\bm{s}}_i\ubar{\bm{s}}_i^* | \F_{i-1}] & \stackrel{(a)}{=} \sum_{\ell = 1}^{N} \sum_{k = 1}^{N} \E\,[ \bm{\pi}_{\ell}(i)\bm{\pi}_{k}(i)] \cdot \E\,[\bm{\mu}_{\ell}(i) \bm{\mu}_{k}(i)] \E\,[ \ubar{\bm{v}}_{\ell,i}({\bm{w}}_{\c,i-1}) \ubar{\bm{v}}_{k,i}^*({\bm{w}}_{\c,i-1}) | \F_{i-1}] \nn \\
& \stackrel{(b)}{=} \sum_{k=1}^{N} (\bar{\pi}_k^2 + c_{\pi,k,k} ) (\bar{\mu}_k^2 + c_{\mu,k,k} ) R_{k,i}({\bm{w}}_{\c,i-1})
\end{align}
where step (a) is by using the independence condition from the asynchronous model in Part I \cite{Zhao13TSPasync1}; and step (b) is from \eqref{I-eqn:Rkidef} in Part I \cite{Zhao13TSPasync1}, \eqref{II-eqn:bigRiandRki} in Part II \cite{Zhao13TSPasync2}, and \eqref{eqn:randstepsizemeanentry}--\eqref{eqn:randcombinecoventry}. Therefore,
\be
\label{eqn:sicovariance2}
\E \ubar{\bm{s}}_i \ubar{\bm{s}}_i^* = \sum_{k=1}^{N} (\bar{\pi}_k^2 + c_{\pi,k,k} ) (\bar{\mu}_k^2 + c_{\mu,k,k} ) \E R_{k,i}({\bm{w}}_{\c,i-1})
\ee
Note that
\begin{align}
\| R_{k,i}(w^o) - \E R_{k,i}({\bm{w}}_{\c,i-1}) \| 
& \stackrel{(a)}{\le} \| \Rcal_i(\one_N \kron w^o) - \E \Rcal_i(\one_N \kron {\bm{w}}_{\c,i-1}) \| \nn \\
& \stackrel{(b)}{\le} \kappa_v \cdot [ \E \, \| \one_N \kron \wt{\bm{w}}_{\c,i-1} \|^4 ]^{\gamma_v/4} \nn \\
& \stackrel{(c)}{=} \kappa_v N^{\gamma_v / 2} \cdot [ \E \, \| \wt{\bm{w}}_{\c,i-1} \|^4 ]^{\gamma_v/4} 
\end{align}
where step (a) is due to \eqref{II-eqn:bigRiandRki} from Part II \cite{Zhao13TSPasync2}; step (b) is by using \eqref{II-eqn:bounddiffRi} also from Part II \cite{Zhao13TSPasync2}; and step (c) is by the fact that $\| \one_N \kron x \|^4 = [ N \cdot \| x \|^2 ]^2 = N^2 \cdot \| x \|^4$ for any $x$. Under Assumption \ref{asm:smallstepsizes}, we can get from Theorem \ref{theorem:4thmomentstability} that 
\begin{align}
\label{eqn:boundgapRkiandRkiw}
\limsup_{i\rightarrow \infty} \| R_{k,i}(w^o) - \E R_{k,i}({\bm{w}}_{\c,i-1}) \| & \le \kappa_v N^{\gamma_v / 2} \cdot [ b_4^2 \nu^2 ]^{\gamma_v/4} = O(\nu^{\gamma_v/2})
\end{align}
which means that, asymptotically, we can replace $\E R_{k,i}({\bm{w}}_{\c,i-1})$ by $R_k$ from \eqref{II-eqn:Rkdef} of Part II \cite{Zhao13TSPasync2} within an error in the order of $\nu^{\gamma_v/2}$. Therefore, it follows from \eqref{eqn:limiyci} that
\begin{align}
\label{eqn:sicovariance3}
\lim_{i \rightarrow \infty} y_{c,i} & = \vecm\left( \lim_{i \rightarrow \infty} \E \ubar{\bm{s}}_i \ubar{\bm{s}}_i^* \right) \nn \\
& \stackrel{(a)}{=} \vecm\left( \sum_{k=1}^{N} (\bar{\pi}_k^2 + c_{\pi,k,k} ) (\bar{\mu}_k^2 + c_{\mu,k,k} ) \lim_{i \rightarrow \infty} \E R_{k,i}({\bm{w}}_{\c,i-1}) \right) \nn \\
& \stackrel{(b)}{=} \vecm\left( \sum_{k=1}^{N} (\bar{\pi}_k^2 + c_{\pi,k,k} ) (\bar{\mu}_k^2 + c_{\mu,k,k} ) [ R_k + O(\nu^{\gamma_v/2}) ] \right) \nn \\
& \stackrel{(c)}{=} \vecm ( R_\c ) + O(\nu^{2 + \gamma_v/2})  
\end{align}
where step (a) is by using \eqref{eqn:sicovariance2}; step (b) is by using \eqref{eqn:boundgapRkiandRkiw}; and step (c) is by using \eqref{eqn:Rcdef} and the fact from \eqref{I-eqn:boundmumoments2} of Part I \cite{Zhao13TSPasync1} that $\bar{\mu}_k^2 + c_{\mu,k,k} = \bar{\mu}_k^{(2)} = O(\nu^2)$. Substituting \eqref{eqn:sicovariance3} into \eqref{eqn:zcinfdef} yields
\be
z_{\c,\infty} = (I_{4M^2}  -  F_\c)^{-1} \cdot \vecm ( R_\c ) + O(\nu^{1 + \gamma_v/2})
\ee
where we used Lemma \ref{lemma:propertiesFc}. Substituting \eqref{eqn:zcinfdef} and $\Sigma = I_{2M}$ into \eqref{eqn:weightedMSEmetric}, and using \eqref{eqn:IminusFcinvOrder} as well as the fact that $F_\c$ and $R_\c$ are Hermitian, we obtain
\begin{align}
\label{eqn:weightedMSEmetric2}
\lim_{i\rightarrow\infty} \E\|\wt{\bm{w}}_{\c,i}'\|^2 & = \frac{1}{2} [\vecm ( R_\c )]^* (I_{4M^2}  -  F_\c)^{-1} \vecm(I_{2M}) + O(\nu^{1 + \gamma_v/2})
\end{align}
Substituting \eqref{eqn:weightedMSEmetric2} into \eqref{eqn:gapMSD} yields \eqref{eqn:MSDbatchexpression}.

We establish \eqref{eqn:MSDbatchapprox} next. From \eqref{eqn:IminusFcexpression}, we know that
\be
\label{eqn:IminusFcequalScOnu2}
I_{4M^2} - F_\c = S_\c + O(\nu^2)
\ee
where
\be
\label{eqn:Scdef}
S_\c \defeq H_\c^\T \kron I_{2M} + I_{2M} \kron H_\c = O(\nu)
\ee
Since $H_\c$ is symmetric and positive definite by \eqref{eqn:eigenvalueHc}, it is easy to verify that $S_\c$ is also symmetric and positive definite. Therefore, $S_\c$ is invertible. Using the matrix inversion lemma \cite{Laub05}, we get from \eqref{eqn:IminusFcequalScOnu2} that
\be
\label{eqn:IminusFcinvScinv}
(I_{4M^2} - F_\c)^{-1} = S_\c^{-1} + O(1)
\ee
where we used the fact that $\| S_\c^{-1} \| = O(
\nu^{-1})$. Substituting \eqref{eqn:IminusFcinvScinv} into \eqref{eqn:MSDbatchexpression} yields:
\begin{align}
\label{eqn:MSDbatchapproxexpress1}
\MSD^{\cent} & = \frac{1}{2} [\vecm ( R_\c )]^* [S_\c^{-1} + O(1)] \vecm(I_{2M}) + O(\nu^{1+\gamma_o}) \nn \\
& = \frac{1}{2} [\vecm ( R_\c )]^* S_\c^{-1} \vecm(I_{2M}) + O(\nu^2) + O(\nu^{1+\gamma_o}) \nn \\
& = \frac{1}{2} [\vecm ( R_\c )]^* S_\c^{-1} \vecm(I_{2M}) + O(\nu^{1+\gamma_o})
\end{align}
where we used the fact from \eqref{eqn:Rcdef} that $\| R_\c \| = O(\nu^2)$ and $\gamma_o < 1/2$ from \eqref{II-eqn:gammaodef} of Part II \cite{Zhao13TSPasync2}. Since the first term on the RHS of \eqref{eqn:MSDbatchapproxexpress1} is of the order of $\nu$, it is the dominant term under Assumption \ref{asm:smallstepsizes}. To further simplify \eqref{eqn:MSDbatchapproxexpress1}, we introduce the Lyapunov equation with respect to the unknown square matrix $X$: 
\be
\label{eqn:Lyapunovequation}
X H_\c + H_\c X = I_{2M}
\ee
where $H_\c$ is given by \eqref{eqn:Hcdef}. Vectorizing both sides and using \eqref{eqn:Scdef}, the Lyapunov equation is equivalent to the linear system of equations:
\be
\label{eqn:linearequation}
S_\c  \vecm(X) = \vecm(I_{2M})
\ee
Since $S_\c$ is invertible, the linear equation \eqref{eqn:linearequation} has a unique solution, which is given by $X = \frac{1}{2} H_\c^{-1}$. From the Lyapunov equation \eqref{eqn:Lyapunovequation} we get
\begin{align}
\label{eqn:ycScinvI}
[\vecm ( R_\c )]^* S_\c^{-1} \vecm(I_{2M}) & = \frac{1}{2} [\vecm ( R_\c )]^* \vecm(H_\c^{-1}) \nn \\
& = \frac{1}{2} \Tr(H_\c^{-1} R_\c)
\end{align}
where we used the fact that $R_\c$ is Hermitian. Result \eqref{eqn:MSDbatchapprox} then follows from \eqref{eqn:MSDbatchapproxexpress1} and \eqref{eqn:ycScinvI}. The term $\Tr(H_\c^{-1} R_\c) = O(\nu)$ in \eqref{eqn:MSDbatchapprox} is the dominant term under Assumption \ref{asm:smallstepsizes}.

\section{Proof of Lemma \ref{lemma:nonnegativity}}
\label{app:nonnegativity}
From Lemma \ref{II-lemma:Ppandp} of Part II \cite{Zhao13TSPasync2}, we know that $P_p$ is symmetric and, therefore, the matrix difference $C_p \defeq P_p - \bar{p}\bar{p}^\T$ is also symmetric. We also know from Lemma \ref{II-lemma:Ppandp} of Part II \cite{Zhao13TSPasync2} that $C_p \one_N = 0$. To establish that $C_p$ is positive semi-definite, we consider the following quadratic expression:
\be
\label{eqn:xpxminusxp2}
x^\T C_p x = x^\T (P_p - \bar{p}\bar{p}^\T) x = x^\T P_p x - (x^\T\bar{p})^2
\ee
for any vector $x \in \mbbR^{N}$. Note that
\be
\label{eqn:xPpx2}
x^\T P_p x = \vecm(x^\T P_p x) = \frac{1}{N^2} (x^\T \kron x^\T) p \cdot \one_{N^2}^\T \one_{N^2}
\ee
by using the relation $p = \vecm(P_p)$ from \eqref{II-eqn:Ppdef} of Part II \cite{Zhao13TSPasync2} and the fact that $\one_{N^2}^\T \one_{N^2} = N^2$. Since 
\be
\label{eqn:AACAandEAA}
\bar{A} \kron \bar{A} + C_A = \E(\bm{A}_j \kron \bm{A}_j)
\ee
we can introduce a series of fictitious random combination matrices $\{\bm{A}_j';j\ge1\}$ such that they are mutually-independent and satisfy 
\be
\E(\bm{A}_j' \kron \bm{A}_j') = \bar{A} \kron \bar{A} + C_A
\ee
for any $j\ge1$. Let $\bm{\Phi}_i \defeq \prod_{j=1}^i \bm{A}_j'$ for any $i\ge1$. Then, 
\begin{align}
\label{eqn:PhikronPhi}
\lim_{i\rightarrow\infty} \E(\bm{\Phi}_i \kron \bm{\Phi}_i) \stackrel{(a)}{=} \lim_{i\rightarrow\infty} \prod_{j=1}^i \E(\bm{A}_j' \kron \bm{A}_j' ) \stackrel{(b)}{=} p \cdot \one_{N^2}^\T
\end{align}
where step (a) is by using the fact that the $\{\bm{A}_j'\}$ are mutually-independent, and step (b) is by using \eqref{eqn:AACAandEAA} and the Perron-Frobenius Theorem \cite{BermanPF}. Substituting \eqref{eqn:PhikronPhi} into \eqref{eqn:xPpx2} and using $\one_{N^2} = \one_N \kron \one_N$, we get
\be
\label{eqn:xPpx3}
x^\T P_p x = \frac{1}{N^2} \lim_{i\rightarrow\infty} \E\,[(x^\T \bm{\Phi}_i \one_N )^2]
\ee
Moreover, since $\bar{A}  =  \E(\bm{A}_j | \bm{w}_{j-1})$, we have
\begin{align}
\label{eqn:Phi}
\lim_{i\rightarrow\infty} \E(\bm{\Phi}_i) = \lim_{i\rightarrow\infty} \prod_{j=1}^i \E(\bm{A}_j') = \lim_{i\rightarrow\infty} (\bar{A})^i = \bar{p} \cdot \one_{N}^\T
\end{align}
Then, using \eqref{eqn:Phi} and the fact that $\one_N^\T \one_N = N$, we have
\be
\label{eqn:xbarp}
x^\T \bar{p} = \frac{1}{N} x^\T \bar{p} \cdot \one_N^\T \one_N = \frac{1}{N} \lim_{i\rightarrow\infty} \E(x^\T \bm{\Phi}_i \one_N) 
\ee
Substituting \eqref{eqn:xPpx3} and \eqref{eqn:xbarp} into \eqref{eqn:xpxminusxp2} yields
\be
x^\T C_p x = \frac{1}{N^2} \lim_{i\rightarrow\infty} \left\{ \E\,[(x^\T \bm{\Phi}_i \one_N )^2]  -  [\E(x^\T \bm{\Phi}_i \one_N)]^2 \right\} \ge 0
\ee
which confirms that $C_p$ is positive semi-definite.

\section{Proof of Lemma \ref{lemma:matchingmeanconvergencerate}}
\label{app:matchingmeanconvergencerate}
We prove Lemma \ref{lemma:nonnegativity} by using a procedure similar to the one given in Appendix \ref{II-app:lowrank} of Part II \cite{Zhao13TSPasync2}. Introduce the Jordan decomposition \cite{Laub05}:
\be
\label{eqn:eigenbarA}
\bar{A} = \bar{P} \bar{J} \bar{Q}^\T = \begin{bmatrix}
\bar{p} & \bar{P}' 
\end{bmatrix} \begin{bmatrix}
1 & 0 \\
0 & \bar{J}' \\
\end{bmatrix} \begin{bmatrix}
\one_N & \bar{Q}'
\end{bmatrix}^\T
\ee
where $\bar{J}'$ is a sub-matrix of $\bar{J}$ containing its stable eigenvalues, $\bar{P}'$ and $\bar{Q}'$ are sub-matrices of $\bar{P}$ and $\bar{Q}$, and $\bar{P}^{-1} = \bar{Q}^\T$. Then, the Jordan decomposition of $\bar{\Acal} = \bar{A} \kron I_{2M}$ from \eqref{II-eqn:bigAmeandef} of Part II \cite{Zhao13TSPasync2} is given by
\be
\label{eqn:JordanbigA}
\bar{\Acal} = \bar{\Pcal} \bar{\Jcal} \bar{\Qcal}^\T = \begin{bmatrix}
\bar{p}' & \bar{\Pcal}' 
\end{bmatrix} \begin{bmatrix}
I_{2M} & 0 \\
0 & \bar{\Jcal}' \\
\end{bmatrix} \begin{bmatrix}
\bar{q}' & \bar{\Qcal}'
\end{bmatrix}^\T
\ee
where
\begin{alignat}{3}
\bar{\Pcal} & = \bar{P} \kron I_{2M}, & \qquad & \bar{\Pcal}' & \defeq \bar{P}' \kron I_{2M} \\
\label{eqn:bigJdef}
\bar{\Jcal} & = \bar{J} \kron I_{2M}, & \qquad & \bar{\Jcal}' & \defeq \bar{J}' \kron I_{2M} \\
\bar{\Qcal} & = \bar{Q} \kron I_{2M}, & \qquad & \bar{\Qcal}' & \defeq \bar{Q}' \kron I_{2M} \\
\label{eqn:pprimeandqprimedef}
\bar{p}' & = \bar{p} \kron I_{2M}, & \qquad & \bar{q}' & \defeq \one_N \kron I_{2M} 
\end{alignat}
Let
\be
\label{eqn:bigXdef}
\bar{\Xcal} \defeq I_{2MN} - \bar{\Dcal} = \bar{\Mcal} \Hcal = O(\nu)
\ee
where $\{\bar{\Dcal}, \bar{\Mcal}, \Hcal\}$ are from \eqref{II-eqn:bigDmeandef}, \eqref{II-eqn:bigMmeandef}, and \eqref{II-eqn:Hkdef} of Part II \cite{Zhao13TSPasync2}, respectively. Then, by \eqref{II-eqn:bigBmeandef} from Part II \cite{Zhao13TSPasync2} and using the fact that $\bar{\Acal}$ is real and $\bar{\Dcal}$ is Hermitian, we get
\be
\label{eqn:eigenvalueofbigB1}
\bar{\Qcal}^\T \bar{\Bcal}^* \bar{\Pcal}  =  \bar{\Qcal}^\T \bar{\Dcal} \bar{\Acal} \bar{\Pcal}  =  \begin{bmatrix}
I_{2M}  -  \bar{q}'^\T \bar{\Xcal} \bar{p}'   &   -\bar{q}'^\T \bar{\Xcal} \bar{\Pcal}' \bar{\Jcal}' \\
-\bar{\Qcal}'^\T \bar{\Xcal} \bar{p}'   &   \bar{\Jcal}'  -  \bar{\Qcal}'^\T \bar{\Xcal} \bar{\Pcal}' \bar{\Jcal}' \\
\end{bmatrix}     
\ee
Using \eqref{eqn:pprimeandqprimedef} and \eqref{eqn:bigXdef} above and \eqref{II-eqn:Hkdef} and \eqref{II-eqn:bigMmeandef} from Part II \cite{Zhao13TSPasync2}, we obtain
\be
\label{eqn:topleftcorner}
\bar{q}'^\T \bar{\Xcal} \bar{p}' = \bar{q}'^\T \bar{\Mcal} \Hcal \bar{p}' = \sum_{k=1}^{N} \bar{p}_k \bar{\mu}_k H_k = H = O(\nu)
\ee
where $H$ is given by \eqref{II-eqn:Hdef} of Part II \cite{Zhao13TSPasync2}. By \eqref{eqn:bigXdef}, we get
\begin{align}
\label{eqn:otherthreeblocks}
\| \bar{q}'^\T \bar{\Xcal} \bar{\Pcal}' \bar{\Jcal}' \| = O(\nu), \quad 
\| \bar{\Qcal}'^\T \bar{\Xcal} \bar{p}'\| = O(\nu), \quad
\| \bar{\Qcal}'^\T \bar{\Xcal} \bar{\Pcal}' \bar{\Jcal}' \| = O(\nu)
\end{align}
Therefore, we get from \eqref{eqn:eigenvalueofbigB1}--\eqref{eqn:otherthreeblocks} that
\be
\label{eqn:eigenvalueofbigB2}
\bar{\Qcal}^\T \bar{\Bcal}^* \bar{\Pcal} = \begin{bmatrix}
\bar{B}_\d & O(\nu) \\
O(\nu) & \bar{\Jcal}' + O(\nu) \\
\end{bmatrix}
\ee
where
\be
\label{eqn:Bddef}
\bar{B}_\d \defeq I_{2M} - H
\ee
is Hermitian. From \eqref{II-eqn:rhoIminusH} of Part II \cite{Zhao13TSPasync2}, we immediately get
\begin{align}
\label{eqn:eigenvalueBd}
\lambda(\bar{B}_\d) & = \lambda(I_{2M} - H) = 1 - O(\nu) > 0 \\
\label{eqn:rhoBddef}
\rho(\bar{B}_\d) & = 1 - \lambda_{\min}(H) = 1 - O(\nu)
\end{align}
for sufficiently small $\nu$ under Assumption \ref{asm:smallstepsizes}. Conjugating both sides of \eqref{eqn:eigenvalueofbigB2} and using the fact that $\bar{B}_\d$ is Hermitian, we get
\be
\label{eqn:bigbarBsdef}
\bar{\Bcal}_s  \defeq  (\bar{\Qcal}^\T \bar{\Bcal}^* \bar{\Pcal})^*  =  \bar{\Pcal}^* \bar{\Bcal} (\bar{\Qcal}^*)^\T  =  \begin{bmatrix}
\bar{B}_\d & O(\nu) \\
O(\nu) & \bar{\Jcal}'^*  +  O(\nu) \\
\end{bmatrix}  
\ee
Since $\bar{\Bcal}_s$ is similar to $\bar{\Bcal}$, they have the same eigenvalues \cite{Laub05}. Since $\bar{B}_d$ is Hermitian, let us introduce its eigenvalue decomposition as
\be
\bar{B}_d = \bar{U} \bar{\Lambda} \bar{U}^*
\ee
where $\bar{U}$ is a $2M \times 2M$ unitary matrix and $\bar{\Lambda}$ is a $2M \times 2M$ diagonal matrix. The $(N-1) \times (N-1)$ matrix $\bar{J}'$, which contains the stable eigenvalues of $\bar{A}$ in \eqref{eqn:eigenbarA}, can be generally expressed as
\be
\label{eqn:Jpdef}
\bar{J}' = \begin{bmatrix}
\bar{\lambda}_{a,2} &  & \bar{T}' \\
 & \ddots & \\
 0 & & \bar{\lambda}_{a,N} \\
\end{bmatrix}
\ee
where $\{\bar{\lambda}_{a,n}\}$ are the eigenvalues of $\bar{A}$ with $\bar{\lambda}_{a,1} = 1$ and $|\bar{\lambda}_{a,n}| < 1$ for all $n = 2,3,\dots, N$. In \eqref{eqn:Jpdef}, the elements in the strictly upper triangular region $\bar{T}'$ are either 1 or 0, which depend on the Jordan blocks in $\bar{J}'$. Using \eqref{eqn:Jpdef} and \eqref{eqn:bigJdef}, we can express the $(2,2)$ block in \eqref{eqn:bigbarBsdef} as
\be
\label{eqn:bigJpdef}
\bar{\Jcal}'^*  +  O(\nu)  =  \begin{bmatrix}
\bar{\lambda}_{a,2}^* I_{2M}  +  O(\nu)   &            &    O(\nu) \\
                &   \ddots   &                 \\
\bar{\Tcal}'^*  +  O(\nu)               &            &   \bar{\lambda}_{a,N}^* I_{2M}  +  O(\nu)  \\
\end{bmatrix}  
\ee
where the elements in the strictly lower triangular region $\bar{\Tcal}'^*$ are either 1 or 0, which depend on the elements of $\bar{T}'$ in \eqref{eqn:Jpdef}. We now apply a similarity transformation to $\bar{\Bcal}_s$ by multiplying
\be
\bar{\Dcal} \defeq \diag\{ \nu^{\epsilon} \bar{U}, \nu^{2\epsilon} I_{2M}, \nu^{3\epsilon}I_{2M}, \dots, \nu^{N\epsilon}I_{2M} \}
\ee
and its inverse $\bar{\Dcal}^{-1}$ on either side of \eqref{eqn:bigbarBsdef}, where $\epsilon = 1/N$. Using \eqref{eqn:bigbarBsdef} and \eqref{eqn:bigJpdef}, we end up with
\begin{align}
\label{eqn:bigbarBsexpress}
\bar{\Dcal} \bar{\Bcal}_s \bar{\Dcal}^{-1} 
& = \left[\begin{array}{c|c}
\bar{\Lambda}              &  O(\nu^{\epsilon}) \\
\hline
 O(\nu^{1+\epsilon})  &   \begin{array}{ccc}
\bar{\lambda}_{a,2}^* I_{2M}   +   O(\nu)    &         &  O(\nu^{\epsilon}) \\
   &    \ddots        &       \\
O(\nu^{\epsilon})    &        &    \bar{\lambda}_{a,N}^* I_{2M}   +   O(\nu)    \\
\end{array} \\
\end{array}  
\right]
\end{align}
From \eqref{eqn:bigbarBsexpress}, we know that all off-diagonal entries of $\bar{\Dcal} \bar{\Bcal}_s \bar{\Dcal}^{-1}$ are \emph{at least} of the order of $\nu^\epsilon$. Therefore, using Gershgorin Theorem \cite[p.~320]{Golub96} under Assumption \ref{asm:smallstepsizes}, and since $\bar{\Bcal}$ and $\bar{\Bcal}_s$ have the same eigenvalues due to similarity, we get
\be
\label{eqn:rhobigBandrhobigYclose}
| \lambda(\bar{\Bcal})  -  \lambda(\bar{B}_d) | \le O(\nu^{1+\epsilon}) \quad \mbox{or} \quad
| \lambda(\bar{\Bcal})  -  \bar{\lambda}_{a,k}^* | \le O(\nu^{\epsilon})
\ee
where $\lambda(\bar{\Bcal})$ denotes the eigenvalue of $\bar{\Bcal}$ and $k = 2, 3, \dots, N$. Result \eqref{eqn:rhobigBandrhobigYclose} implies that the eigenvalues of $\bar{\Bcal}$ are either located in the Gershgorin circles that are centered at the eigenvalues of $\bar{B}_d$ with radii $O(\nu^{1+\epsilon})$ or in the Gershgorin circles that are centered at $\{\bar{\lambda}_{a,k}^*; k=2,3,\dots,N\}$ with radii $O(\nu^{\epsilon})$. From \eqref{eqn:rhoBddef}, we have
\be
\label{eqn:rhoBdless1}
\rho(\bar{B}_d) = 1 - O(\nu) < 1
\ee
By Assumption \ref{II-asm:connected} from Part II \cite{Zhao13TSPasync2} and Perron-Frobenius Theorem \cite{BermanPF}, we have
\be
\label{eqn:rhorelation2}
\rho(\bar{J}'^*) \defeq \max_{k=2,3,\dots,N} | \bar{\lambda}_{a,k}^* | = \rho(\bar{J}') < 1
\ee
By Assumption \ref{asm:smallstepsizes}, if the parameter $\nu$ is small enough such that
\be
\label{eqn:lambdaBgerhoJp}
\rho(\bar{J}') + O(\nu^\epsilon) < 1 - O(\nu) = \rho(\bar{B}_d)
\ee
holds, then the Gershgorin circles centered at the eigenvalues of $\bar{B}_d$ are isolated from those centered at $\{\bar{\lambda}_{a,k}^*; k=2,3,\dots,N\}$. According to Gershgorin Theorem \cite[p.~181]{Stewart90}, there are precisely $2M$ eigenvalues of $\bar{\Bcal}$ satisfying
\be
\label{eqn:rhobigFclosetolambdaF}
| \lambda(\bar{\Bcal}) - \lambda(\bar{B}_d) | \le O(\nu^{1+\epsilon})
\ee
while all the other eigenvalues satisfy
\be
\label{eqn:rhobigFclosetolambdaak}
| \lambda(\bar{\Bcal}) - \bar{\lambda}_{a,k}^* | \le O(\nu^{\epsilon}), \quad k = 2, 3, \dots, N
\ee
By \eqref{eqn:lambdaBgerhoJp}, the eigenvalues $\lambda(\bar{\Bcal})$ satisfying \eqref{eqn:rhobigFclosetolambdaF} are greater than those satisfying \eqref{eqn:rhobigFclosetolambdaak} in magnitude. Furthermore, when $\nu$ is sufficiently small, the Gershgorin circles centered at $\lambda_{\max}(\bar{B}_d)$ with radius $O(\nu^{1+\epsilon})$ will become disjoint from the other circles. Then, by using Gershgorin Theorem again, we conclude from \eqref{eqn:rhobigFclosetolambdaF} that
\be
\label{eqn:rhobigBandrhoBd}
|\rho(\bar{\Bcal}) - \rho(\bar{B}_d)| \le O(\nu^{1+\epsilon})
\ee
It is worth noting that from \eqref{eqn:rhoBdless1} and \eqref{eqn:rhobigBandrhoBd} we get
\be
\rho(\bar{\Bcal}) \le 1 - O(\nu) + O(\nu^{1+\epsilon}) < 1
\ee
for $\nu \ll 1$ because $\epsilon = 1/N > 1$. Eventually, using \eqref{eqn:Bmeandef}, \eqref{eqn:piandp}, and \eqref{eqn:Bddef}, it is straightforward to verify that
\be
\label{eqn:BandBd}
\bar{B} = \bar{B}_\d
\ee
Using \eqref{eqn:rhoBddef}, \eqref{eqn:rhobigBandrhoBd}, and \eqref{eqn:BandBd} completes the proof.

\end{document}